\documentclass[preprint,5p,times,twocolumn,authoryear]{elsarticle}
\usepackage{amssymb}
\usepackage{amsmath}
\usepackage{array}
\usepackage{lineno}
\usepackage{fontawesome}
\usepackage{wrapfig}
\usepackage{caption}
\usepackage[rightcaption]{sidecap}
\usepackage[most]{tcolorbox}
\usepackage[export]{adjustbox}
\usepackage{dirtytalk}
\usepackage{subcaption}
\usepackage{wrapfig}

\journal{The Journal of Systems and Software}
\begin{document}

\begin{frontmatter}


%%\tnotetext[label1]{}
\author[label0]{Ulrike M. Graetsch} \corref{cor1} %\fnref{label1} %\corref{name}
\ead{ulrike.graetsch@monash.edu}
\author[label0]{Rashina Hoda}
\ead{rashina.hoda@monash.edu}
\author[label1]{Hourieh Khalazjadeh}
\ead{hkhalajzadeh@deakin.edu.au}
\author[label2]{Mojtaba Shahin}
\ead{mojtaba.shahin@rmit.edu.au}
\author[label0]{John Grundy}
\ead{john.grundy@monash.edu}

%\fntext[label2]{}
\cortext[cor1]{Corresponding author}
\affiliation[label0]{organization={Faculty of Information Technology, Monash University},
addressline={Wellington Road}, 
           city={CLAYTON},
           postcode={3800}, 
            state={Victoria},
            country={Australia}}
%\fntext[label3]{}

\title{Managing Technical Debt in a Multidisciplinary Data Intensive Software Team: an Observational Case Study} %% Article title

 \affiliation[label1]{organization={School of Information Technology, Deakin University},
          addressline={221 Burwood Highway},
            city={BURWOOD},
            postcode={3125},
            state={Victoria},
             country={Australia}}
 \affiliation[label2]{organization={School of Computing Technologies, RMIT University},
            addressline={124 La Trobe Street},
             city={MELBOURNE},
             postcode={3000},
             state={Victoria},
             country={Australia}}

%% Abstract
\begin{abstract}
%NOTE - 200 word limit.
\textbf{Context:} There is an increase in the investment and development of data-intensive (DI) solutions – systems that 
manage large amounts of data. Without careful management, this growing investment will also grow associated technical debt (TD). Delivery of DI solutions requires a multidisciplinary skill set, but there is limited knowledge about how multidisciplinary teams develop DI systems and manage TD.\\
\textbf{Objective:} This research contributes empirical, practice based insights about multidisciplinary DI team TD management practices.\\
\textbf{Method:} This research was conducted as an exploratory observation case study. We used socio-technical grounded theory (STGT) \textit{for data analysis} to develop concepts and categories that articulate TD and TDs debt management practices.\\
\textbf{Results:} We identify TD that the DI team deals with, in particular technical data components debt and pipeline debt. We explain how the team manages the TD, assesses TD, what TD treatments they consider and how they implement TD treatments to fit sprint capacity constraints.\\
\textbf{Conclusion:} We align our findings to existing TD and TDM taxonomies, discuss their implications and highlight the need for new implementation patterns and tool support for multidisciplinary DI teams.\\

\end{abstract}


%%\item Research highlight 1
%%\item Research highlight 2

%% Keywords

\end{frontmatter}

%%\linenumbers


\section{Introduction}
\label{sec1}
%% Labels are used to cross-reference an item using \ref command.
Data intensive (DI) solutions are systems that analyse and manipulate data to provide predictions and insights. These systems are becoming pervasive. The big data and business analytics market was valued at US\$215.7B in 2021 \citep{AIMagazine2021} and is projected to be \$924B by 2032 \citep{Fortune2023}. Delivery of DI solutions involves expertise and skills from multidisciplinary teams \citep{Graetsch2023}. DI systems rely on data, not just algorithms or programs to deliver an outcome or result. Examples of DI systems include enterprise data analytics solutions, imaging diagnostic systems, and real estate price predictors. The development of DI systems requires the combination of different disciplines including substantial domain knowledge, software engineering, data science and cloud computing. 
Traditionally, teams are characterised as multidisciplinary teams because team members present with different bodies of knowledge, research communities, ways of working, education and career pathways. \citep{med2006multidisciplinarity}. Within DI systems delivery, cross-skilling between team members of different disciplines is valued, but difficult to achieve \citep{Graetsch2023}. There is growing recognition of the critical role these teams will play in the success of future digital transformation \citep{Gartner2022}, yet there is limited empirical research in the software engineering domain with focus on these teams, their team members or their ways of working. Research to build this understanding can be challenging.  It involves people -- often professional practitioners and dealing with organisations that have goals other than research, and  requires adherence to ethics processes. Consequently, it is carried out less frequently than other empirical research \citep{Storey2020-fh}.\par
In parallel with the explosive DI systems growth, there is a growing awareness of the potential for the associated technical debt to grow. Technical debt \textcolor{black}{(TD)}, first articulated by Cunningham in 1992 \citep{Cunningham1992-mf} is a metaphor for describing the `debt' or effort that is incurred when shortcuts to quality are taken to achieve delivery goals. \textcolor{black}{(TD)} has been a mainstay in software engineering for some time.  \textcolor{black}{TD was defined as referring to “delayed tasks and immature artifacts that constitute a \textcolor{black}{`debt'} because they incur a cost in the future in the form of increased costs of change during evolution and maintenance.”, at the Daghstuhl seminar on Managing Technical Debt in Software Engineering \citep{Avgeriou2016-vj}.}
However, the extension of \textcolor{black}{(TD) metaphor} to DI systems has been much slower -- with the first references by Weber et al. \citep{Weber2014-hz} and \textcolor{black}{Aniche} et al.\citep{Aniche2014-nf} in 2014.  A body of research is slowly building including work to characterise and define DI \textcolor{black}{TD} \citep{Weber2014-hz, Foidl2019, Sculley2015-hq}, measuring DI \textcolor{black}{TD} \citep{sklavenitis2024} and managing DI debt \citep{AlBarak2016, Albarak2018-jy, Albarak2020-vd, Muse2022, Muse2022-aa, Kleinwaks2023-xi, Waltersdorfer2020-yi}.\par
\textcolor{black}{
Empirical studies about TD and TD Management in software engineering have considered perspectives of practitioners \citep{Perez2021-xm, Li2023-zc, FREIRE2023, Guo2016-hc, Xavier2023-fs}. Empirical research about TD in DI systems is an active area of research  \citep{Tang2021-dr, Muse2022, Albarak2020-vd, Waltersdorfer2020-yi}. However there is limited research on practitioners’ perspectives \citep{Recupito2024-gw}. Empirical studies of TD in DI systems have mainly involved developers\citep{Aniche2014-nf, Weber2014-hz, Albarak2020-vd}, but a wider range of practitioners, or multidisciplinary teams have not yet been studied. How TD and TD Management research aligns with day to day work practices of multidisciplinary teams that develop DI systems remains underexplored.} 
\par This research builds knowledge to contribute and extend knowledge about DI \textcolor{black}{TD} and its management and builds on our prior research into the challenges that multidisciplinary DI teams face and how they deal with data challenges \citep{Graetsch2023}. We aim to deepen understanding of multidisciplinary DI team data practices through detailed observations, including their practices relating to \textcolor{black}{TD}. We designed an exploratory observation case study to observe a 12 member agile data analytics delivery team in a large organisation. We observed the team for 6 weeks and collected data for analysis, including rich descriptions of the case context. We applied socio-technical grounded theory \textcolor{black}{\textit{for data analysis} \citep[Chapter 10]{Hoda24}} to develop concepts and categories about \textcolor{black}{TD} and its management and present these in our findings. We discuss the findings in the context of related literature. Key findings of the study include details about the \textcolor{black}{TD} that DI teams discuss, and the team’s practices such as assessing \textcolor{black}{TD} and structuring work to address \textcolor{black}{TD} during %agile
Scrum ceremonies. The team balances their desire for holistic solutions with the reality of limited capacity, using splitting strategies to achieve treatment of \textcolor{black}{TD} so they can deliver within the boundaries of their sprint capacity. We highlight the need for new patterns to pay down or treat \textcolor{black}{TD} that can be adapted to multidisciplinary DI system delivery.\par
This research makes the following key contributions: 
\begin{itemize}
    \item We present a detailed observational case study that provides descriptions and contextual information about multidisciplinary team member work practices in an agile data-analytics team. 
    \item We conceptualise technical debt related discussions in a multidisciplinary data-analytics team, including the contexts in which technical debt is identified and assessed, how it assessed and  treatment approaches considered.
    \item We provide a set of recommendations for practitioners and key future work directions for researchers.
\end{itemize}  
\par
The rest of this paper is organised as follows: We present related works in Section 2. Section 3 provides an overview of our research method. We present our study findings in Section 4 and discuss our findings in terms of the related works in Section 5 where we also present key implications and recommendations for both practitioners and researchers.  Section 6 considers the threats and limitations of our study and Section 7 provides a concluding summary.  
\vspace{.5cm}

\section{Background and Motivation} \label {sec:motivation}

\subsection{Technical Debt and Technical Debt Management} 
Technical debt \textcolor{black}{(TD)}, first articulated by Cunningham in 1992 \citep{Cunningham1992-mf} is a metaphor for describing the `debt' or effort that is incurred when shortcuts to quality are taken to achieve delivery goals. The resulting growth in empirical research led to systematic reviews and early conceptualisations of \textcolor{black}{TD} \citep{Li2015-vm, Tom2013-vo}. The Dagstuhl Seminar 16162 \citep{Avgeriou2016-vj} developed a research roadmap to develop two view points of describing \textcolor{black}{TD} for software-intensive systems; firstly to explore \textcolor{black}{TD} properties, artifacts and elements, and secondly, to explore the management of \textcolor{black}{TD} encompassing the process related activities or different states the debt may go through \citep{Avgeriou2016-vj}. %At that seminar, TD management was described as recognizing, analyzing, monitoring, and measuring Technical Debt.
\par
\textcolor{black}{TD} and \textcolor{black}{TD} management have been systematically analysed from the perspective of traditional software engineering \citep{Li2015-vm, Alves2016-no, Rios2018-yi, Tom2013-vo}, with Rios et al. synthesising the prior research through their tertiary study \citep{Rios2018-yi}.  They also assessed activities, tools and strategies to support \textcolor{black}{TD} management activities and categorised these into 4 macro TDM activities: prevention, identification, monitoring, and payment.
A case study by \cite{Guo2016-hc} studied the effects of implementing a \textcolor{black}{TD} management approach to better understand the cost and benefit of explicitly managing \textcolor{black}{TD}, to contribute to the development of a theory about \textcolor{black}{TD} management \citep{SEAMAN201125}. \textcolor{black}{Perez et al. conducted a study to empirically explore the practices used for TD Payment and TD Prevention and \textcolor{black} {cataloged} a set of TD practices \textcolor{black}{that can prevent or reduce TD from the software architects' point of view}  \citep{Perez2021-xm}}. More recently, \cite{FREIRE2023} conducted a survey study of software practitioners \textcolor{black}{to gain} insight into \textcolor{black}{TD} payment methods and reasons for not repaying \textcolor{black}{TD}. They identified that \textcolor{black}{TD} management is complex and involves a combination of managerial and technical practices, which makes it important for managers and technical team members to be involved in \textcolor{black}{TD} repayment \citep{FREIRE2023}. \textcolor{black}{They also created a TD Payment map, which extends Rios et al.'s TDM Landscape \citep{FREIRE2023}. The creation and management of Self Admitted TD (SATD) in an industry setting was studied by Li et al. in their exploratory case study\citep{Li2023-zc}. Their study investigated characteristics of SATD in different sources, including commit messages, code comments, and issue trackers.  Most relevant to our research, they considered developer perspectives to identify common practices used to support management and pay back of SATD, as well as feature requirements for tools to manage SATD.}

The foundational works and more recent works were focused on software intensive systems and did not include \textcolor{black}{TD} considerations for data-intensive systems.

\subsection{Data-intensive Software System Technical Debt}
Two of the earliest data related \textcolor{black}{TD} challenges were about missing referential integrity constraints at the database level \citep{Weber2014-hz} and placing Data Access Object methods into the wrong class \citep{Aniche2014-nf}. The articulation of specific technical data debt was followed by development of a taxonomy of database design related technical debt   \citep{AlBarak2016}. 
\par
Sculley et al. raised the concept of specialised technical debt in machine learning systems, specifically due to their data-intensive nature. They identified debt at the system level, as well as data debt in machine learning systems \citep{Sculley2015-hq}. A conceptual model to articulate where DI \textcolor{black}{TD} can emerge was articulated by Foidl et al. \citep{Foidl2019}. The model shows that within a Data Intensive Software System (DISS), \textcolor{black}{TD} emerges due to software architecture or software implementation decisions, as well as Data Model or Data Storage constructs, and/or through data - specifically data quality issues \citep{Foidl2019}. One important implication of including data quality as a factor is that the \textcolor{black}{TD} of 2 implementations of the same DISS can be different, if their data is different, and identifying and treatment of \textcolor{black}{TD} in DISS needs to consider the underlying data and its criticality \citep{Weber2014-hz}.\par
Muse et al. investigated data access related \textcolor{black}{TD} and performance anti-patterns in a selection of SQL and NoSQL databases \citep{Muse2022, Muse2022-aa}. They investigated and categorised the self admitted technical debt (SATD) in data access related source code commits and proposed an `data access' extension to Albarak and Bahsoon's taxonomy and downstream \textcolor{black}{TD} identification tools \citep{Muse2022}. 
\par
The systems engineering discipline is also only beginning to apply and research the \textcolor{black}{TD} metaphor \citep{Kleinwaks2023-xi}. The need to address data model debt in the systems engineering discipline was identified by Waltersdorfer et al. who developed a Production System framework that identified causes and mitigation approaches for new \textcolor{black}{TD} types. They incorporated a multidisciplinary perspective and developed a framework covering data model debt, knowledge representation debt and exchange process debt \citep{Waltersdorfer2020-yi}. 

\subsection{Data-Intensive Technical Debt Management}
\par
Research on managing data intensive \textcolor{black}{TD} and how it differs from traditional technical debt management is still emerging. Muse et al. considered the time it takes to address SATD in data-intensive systems compared to traditional technical debt, and found that SATD remains active twice as long compared to traditional technical debt. They also found that bug fixing and refactoring were the main reasons for introducing SATD and that removal of data access SATDs was mostly associated with feature enhancements, new feature introduction and bug fixing, but not refactoring \citep{Muse2022}. Tang et al. studied refactoring in open-source ML systems and created a Hierarchical ML specific refactoring taxonomy \citep{Tang2021-dr}. The need to consider and balance data quality, refactoring and associated data migration costs motivated Albarak et al.'s development of a multi attribute decision analysis framework for prioritising database normalisation debt \citep{Albarak2020-vd, Albarak2018-jy}. Sklavenitis and Kalles have proposed a methodology for quantification \textcolor{black}{TD} within competitive AI platforms. They consolidated and categorised existing \textcolor{black}{TD} research studies into \textcolor{black}{TD} types and factors relevant for measuring the technical debt type.  Where possible they also identified mitigating strategies. Their methodology is under evaluation  \citep{sklavenitis2024}.

\subsection{Consideration of Practitioner Perspectives}
Whilst empirical research about \textcolor{black}{TD} in DI systems is an active area of research \citep{Tang2021-dr, Muse2022, Muse2022-aa, Albarak2020-vd, Waltersdorfer2020-yi, Recupito2024-gw}, there is limited research on the practitioner perspective \citep{Recupito2024-gw}. Some studies involved evaluations with developers \citep{Aniche2014-nf, Weber2014-hz, Albarak2020-vd}, or wider set of participants \citep{Waltersdorfer2020-yi}, whereas Recupito et al. provide the first insights into state of the art practitioner perspectives on technical debt in the context of artificial intelligence systems (AITD). They selected a subset of hidden \textcolor{black}{TD} scenarios first articulated by Sculley at al. \citep{Sculley2015-hq} nearly 10 years ago. Surprisingly, Recupito et al. found a low prevalence of AITD issues identified by practitioners, indicating that the state of practice of \textcolor{black}{TD} management for AI enabled systems is at a preliminary state and participants are not able to recognise issues in their systems \citep{Recupito2024-gw}. Practitioners perspectives have been considered in software specific \textcolor{black}{TD} management practices \textcolor{black}{\citep{FREIRE2023, Guo2016-hc, Xavier2023-fs, Li2023-zc, Perez2021-xm}} but the extension of this work
to multidisciplinary data-intensive software teams remains under researched.

\subsection{Motivation for Our Study}

Our previous exploratory interview study developed a theory about multidisciplinary data-intensive development teams having to deal with data challenges and identified strategies and approaches that teams use to address these 
 \citep{Graetsch2023}. Whilst the findings of that study motivate the need for and chart directions for possible solutions, they lack detail about the actual work context on the practices of multidisciplinary data-intensive teams.  The study did not capture  (i) how key data-related challenge issues emerge in data-intensive software, (ii) how multidisciplinary teams identify, evaluate and resolve such issues, and (iii) what role their respective development tools and workflows play during team member interactions. 
 
 We wanted to take a human-centred approach to designing and developing solutions, but had insufficient information to drive use cases for solution development. A wide search of the literature about data practices/tools/multidisciplinary teams did not yield adequate detailed information. 
To address this, we developed a research agenda with the aim of filling this current gap in knowledge about multidisciplinary, data-intensive system teams and how they work, with a human-centred lens to focus on data work practices to:\\
\begin{itemize}
    \item Contextualise and drive understanding of data-intensive delivery work practices
    \item Understand how experts in a multidisciplinary team work together on data scenarios
    \item Understand how they use their tools when working together
\end{itemize}

To achieve this aim we decided to adopt a fieldwork based case study approach in the form of a participant observation study \textcolor{black}{\citep{Runeson2012}} using socio-technical grounded theory \textcolor{black}{for data} analysis \citep{Hoda24} (See Section 3). We chose this approach over surveys and interviews to gain a greater depth of understanding of how a team works together on a project than is achievable through interviews or repository mining studies. Whilst conducting this 6 week observational field study, it became apparent that there were patterns of discussions and work by the team focused on identifying and managing data-related \textcolor{black}{TD}. Through our analysis, it became clear that this data-related \textcolor{black}{TD} was an important category of work done by the team -- a sizable portion\textcolor{black}{, i.e., }more than 30\% of the observed interactions referred to or discussed one or more \textcolor{black}{TD} or \textcolor{black}{TD} management concepts.  We therefore wanted, in this paper, to analyse this large observational field study dataset to answer the following key research questions around the team's discussions and interactions about technical debt:\\
\begin{itemize}
    \item RQ1: What does a multidisciplinary, data-intensive system engineering team discuss about technical debt?
    \item RQ2: How does the team identify and assess technical debt?
    \item RQ3: What does such as team discuss about technical debt treatment?
    \item RQ4: How does the team decide the treatment it will apply to the technical debt?
\end {itemize} 

\section{Research Method} 
\subsection{Software Engineering Case Study}
We selected the Case study research method as it supports empirical enquiry in real-life contexts; in particular, this method enables researchers to draw on multiple sources of evidence to investigate an instance of a ``contemporary software engineering phenomenon within its real-life context, especially when the boundary between phenomenon and context cannot be clearly speciﬁed" \citep{Runeson2012}. 

Runeson et al.'s Case Study method and guidelines have been specifically targeted to software engineering \citep{Runeson2012, Wohlin2021-zo}. More recently, Wohlin suggested refinements to the definition of case study to focus on using multiple data collection methods, studying a contemporary (not historical) phenomenon in real-life (with people) context, with the investigator not taking an active role \citep{Wohlin2021-zo}.  Case Studies are most suited to answering questions of an exploratory nature and need to accept that the researcher has a low level of control over the study situation, in return for realism \citep{Runeson2012}. \par  

Our aim was to study the phenomenon of a multidisciplinary data-intensive software team in situ, seeking a strong element of realism and detail, and the researcher was taking an observational role and not taking an active role in the study. Our research aim also aligned with Wohlin's more nuanced definition as we sought to investigate and develop an understanding of the practitioners' perspective and work practices\textcolor{black}{, i.e.,} how a multidisciplinary team of experts work together in a data-intensive software team.    
\subsection{Case Study Design} \label {subsec4}
We followed the guidance of Runeson et al. regarding the elements of case research design \textcolor{black}{\citep{Runeson2012}}. Based on the rationale and purpose set out in Section 2 above, we defined the object, or case, to be studied as an ``exploration of a team of professional practitioners delivering a data-intensive solution", specifically the observable practices of the team members and how they \textcolor{black}{interrelate}. This requires a broad perspective and as such we determined that a holistic design, where the case is the unit of analysis, would be most appropriate. Under this design, the context of the case consists of both the organisational context and the coordination\textcolor{black}{, i.e.,} Scrum ceremonies.\par
The case study was designed such that the researcher should take an observational role only, with participants being fully informed and aware of the observations at all times. The researcher was seen as having the role of a researcher, and was not part of the delivery team. The aim was to observe the team members without interrupting their normal flow and to observe what actually happens.\par
The ceremonies were key units of observation\textcolor{black}{, i.e.,} the researcher planned to attend as many team ceremonies as possible during the duration of the case study to observe team interactions. The design of our timeline had to balance our desire to observe end to end delivery, provide opportunity to observe reflective team meetings resulting in changes to practices \citep{Dittrich2020} and respect the commercial realities of operating a team under a research environment. We estimated a timeline of 6 weeks based on the assumption that if the observed team worked in an agile, \textcolor{black}{specifically, Scrum} delivery environment with 2-week sprints, it would allow observation of 3 sprints to provide observation opportunities for end to end delivery of features as well as provide insights into team reflection, continuous improvement efforts and in addition, the first author also planned to attend deep dive sessions to observe detailed interactions between team members. Whilst we planned to take an observer role only, we also planned for questions that may arise during observations about the context and practices and hence included clarification sessions with individual team members where we could ask questions in our case protocol.\par
Whilst observations were the main data source, we also planned to collect data about work items that the team members worked on, and interactions or messages that the team members exchanged. The case was designed for this data to be collected through observation notes, video recording interactions and access to messages being connected to the team collaboration system.     
To prepare and plan the observations, the first author reviewed literature about ethnographic observations \citep{sharp2004ethnographic, Sharp2016, SHAH2014, fetterman2019ethnography} and created templates and procedures for data collection \citep{spradley2016participant}.  She also completed self-paced observation exercises that focus on different aspects of observations to improve observation skills\citep{nippert2015watching}.

\subsubsection{Human Ethics, Privacy and Confidentiality}
Whilst the need for more field work based case studies with human subjects in Software Engineering has been raised \citep{Storey2020-fh}, these studies come with additional requirements regarding data collection and management. Entering the natural settings of participants, for a lengthy period of time, requires careful consideration of how to manage the interactions and any obligations that arise, and how to ensure reciprocity \citep{Hammersley2020-ns}. These considerations need to be addressed throughout the research study, but come to fore first during the ethics process. We followed the guidelines by Runeson et al. \citep{Runeson2012} and our planned case study protocol was reviewed and approved by the Monash University Human Research and Ethics Committee (MUHREC).\par
 
The involvement of practitioners working in an organisation carries the additional requirement to protect commercially sensitive information. As we planned to video record and transcribe observations, we developed a protocol to collect data using the organisation's Zoom platform, and transfer the recordings to our university Google Drive. We committed to `blurring' faces in the video recordings to protect participants' privacy, and blurring commercially sensitive data or information. We planned to have the voice recordings transcribed using Otter.ai and committed to removing the recordings from the platform at completion. We also committed to anonymising the Otter.ai transcripts. At conclusion of the data collection phase, all data would be stored in our secure university Google Drive, and our stored video recordings were blurred and our transcripts anonymised.\par
We developed recruitment fliers and emails. To support the requirement for `informed consent', we prepared materials for different types of audiences. The case study protocol, including the data management protocol were summarised into a general `recruitment presentation' and then more formally into separate Explanatory Statements for the organisation, team member participants and stakeholders that may attend meetings. We developed a pro-forma permission letter for the organisation, consent forms for team members and consent forms for stakeholders. We also created a ``Zoom background" slide to be used by the researcher when attending Zoom calls. The background slide served to alert attendees that an observation was in progress and provided a QR code with links to the explanatory materials. We also created a `flyer' for meeting room doors to inform any attendees who physically attend that the observation study was being conducted. We also committed to verbally notifying attendees that a recording would be taking place and seeking consent from all attendees prior to each meeting.\par
All materials were reviewed and approved by MUHREC. The aim of these materials and communications was to ensure that any attendee at the meeting was informed about the study and had the opportunity to provide consent prior to each observation and to request stopping the recording at any point. Further, our process had to ensure that we collected  one formally completed consent form from each participant (team member or stakeholder) who participated in one or more observation meetings. We make these materials available for reference in supplementary materials\footnote{ https://doi.org/10.5281/zenodo.13377355}.\par
\subsubsection{Recruitment}
The authors posted advertisements on LinkedIn as we wanted to attract professional industry participants, and also reached out to professional networks. \textcolor{black}{The recruitment materials targeted representatives from organisations with teams that met the following high level selection criteria:
1) The team should be multidisciplinary with one or more domain, business or subject matter expert, one or more software experts and one or more data experts. 2) The team should be developing data analytics software, possibly including, machine learning, artificial intelligence or other data analytics features. 3) There should be multiple data sources, including sources from another team, organisation or system\textcolor{black}{, i.e.,} the team should not be generating the data.}

We received \textcolor{black}{three} inquiries and proceeded to present our \textcolor{black}{case study research proposal, which articulated additional criteria about the nature of the project, the nature of the observations and consent requirements: 1) The project should have at least 1-2 months left on its timeline to allow time for 6 weeks observations. 2) The team should embed the researcher \say{into the team} to conduct observations of day to day work, which would include being invited to day to day team meetings or ceremonies and additional activities and exchanges relating to dealing with data challenges. This also included access to relevant team messaging channels/emails and relevant artifacts. 3) The organisation would need to provide written permission from an authorised representative allowing us to conduct the case study and establish contact with team members. 4) Each team member in the team would be willing to provide voluntary, informed consent to participate in the study. Participation would include participation in a focus group and reflective interviews, permission of video recording and photos. Note: participants and the organisation had the ability to withdraw their consent at any stage, until data was anonymised. We make the recruitment materials and explanatory statement for participants available as supplementary materials \footnote{https://doi.org/10.5281/zenodo.13377355}}.

\par \textcolor{black}{Discussions with two of the contact points did not progress past the initial recruitment meeting as they were unable to engage the relevant contact point in their organisation that had the authority to provide us permission to conduct the case study. Discussions with the third contact point progressed and led to engagement and further meetings with the executive level of an organisation.}

% overview slides, of which one proceeded to engagement with the executive level of an organisation. 
%Shortly after, a professional contact of the fifth author expressed interest and the first author presented the overview presentation, highlighting the potential value of the project to his organisation - including the power of reflection, helping build knowledge in the software engineering community and a customised technical report to highlight findings. Whilst the contact person was not authorised to provide permission for their organisation, they proceeded to engage with an executive at their employer who had authority.\par
The first author was invited to present the study proposal to Data Capability Leads at the organisation. The Capability Leads then organised for the first author to present the study overview to the Date Engineering Group (DEG) executive committee. The executives had questions mainly pertaining to confidentiality and privacy and the first author articulated the elements in the protocol pertaining to privacy including anonymisation and data storage protocol. \textcolor{black}{After these sessions, the first author was granted in principle permission to proceed with the case study and to contact the team.} Subsequently, the first author was referred to the `Star Squad'\footnote{Team and organisation name anonymousised at request} project leaders - the Iteration Manager and Product Manager to present the study overview and discuss the study and protocol in more detail \textcolor{black}{specifically the logistics of the observations and how the criteria would be met in detail and to ensure that the observations would not interfere with their day to day work, assuming that team members would consent to the study.}  The final recruitment meeting was with the team members, who were presented with the case study research proposal and the explanatory statements seeking their consent.
\par
\subsubsection{Obtaining Consent}
The project leaders raised the proposed study with team members without the presence of anybody from the research team to see whether the team members would be interested and whether they had any objections. The team members were interested in finding out more. The first author was then invited to present the overview of the study to the project team and provided the team members with copies of the Explanatory Statement and Consent Forms forwarded through the Iteration Manager. The team members had a number of questions about confidentiality and privacy - which were addressed through the data protocol and regular consent processes.\par
The first author was only provided with the names of team members and contact details as part of their completed consent form. Once all team members had provided their consent, the observation study was set to commence on January 30, 2024. The total recruitment duration for this study, from advertisement to commencement was 5 months. The duration from first contact with the organisation representative, including the various presentations and gathering of consent, until study commencement, was 3 months.\par
During the study, there were a number of observation meetings involving stakeholders. The sessions were recorded on Zoom as part of normal proceedings (as it was the Sprint Review), and once stakeholders had provided their formal consent, the recording was released to the first author.

\begin{table}[h]%% placement specifier
\footnotesize
\caption{\textcolor{black}{\\Summary of observation data}}
\vspace{-0.3cm}
  \begin{tabular}{l r r}
  \hline
\textbf{Meeting Type} & \textbf{Observations} & \textbf{Duration (hrs)}\\
 \hline %\hline
 Team Kick Off&1& 3.0 \\
 Sprint Planning&4 & 10.0\\
 Backlog Refinement&5& 6.0\\
 Daily StandUp&29& 9.0 \\
 After Party&18& 6.0 \\
 Deep Dive Session&11& 5.5\\
 Sprint Review&4& 3.5\\
 Sprint Retro&4&  4.0\\
% Clarification Session&15& 7.5\\
 Reflective Workshop&1& 1.0\\
 UAT Workshop &1 &1.0 \\
  \hline
%  \hline
%   Total &93 &56.0 \\
    Total &\textcolor{black}{78}& \textcolor{black}{49.5} \\
   \hline
\end {tabular}
%\caption{Summary of observation data}
\end {table}
\label{tbl:observations}

\begin{figure*} [h]
\includegraphics[width=0.9\textwidth, center]{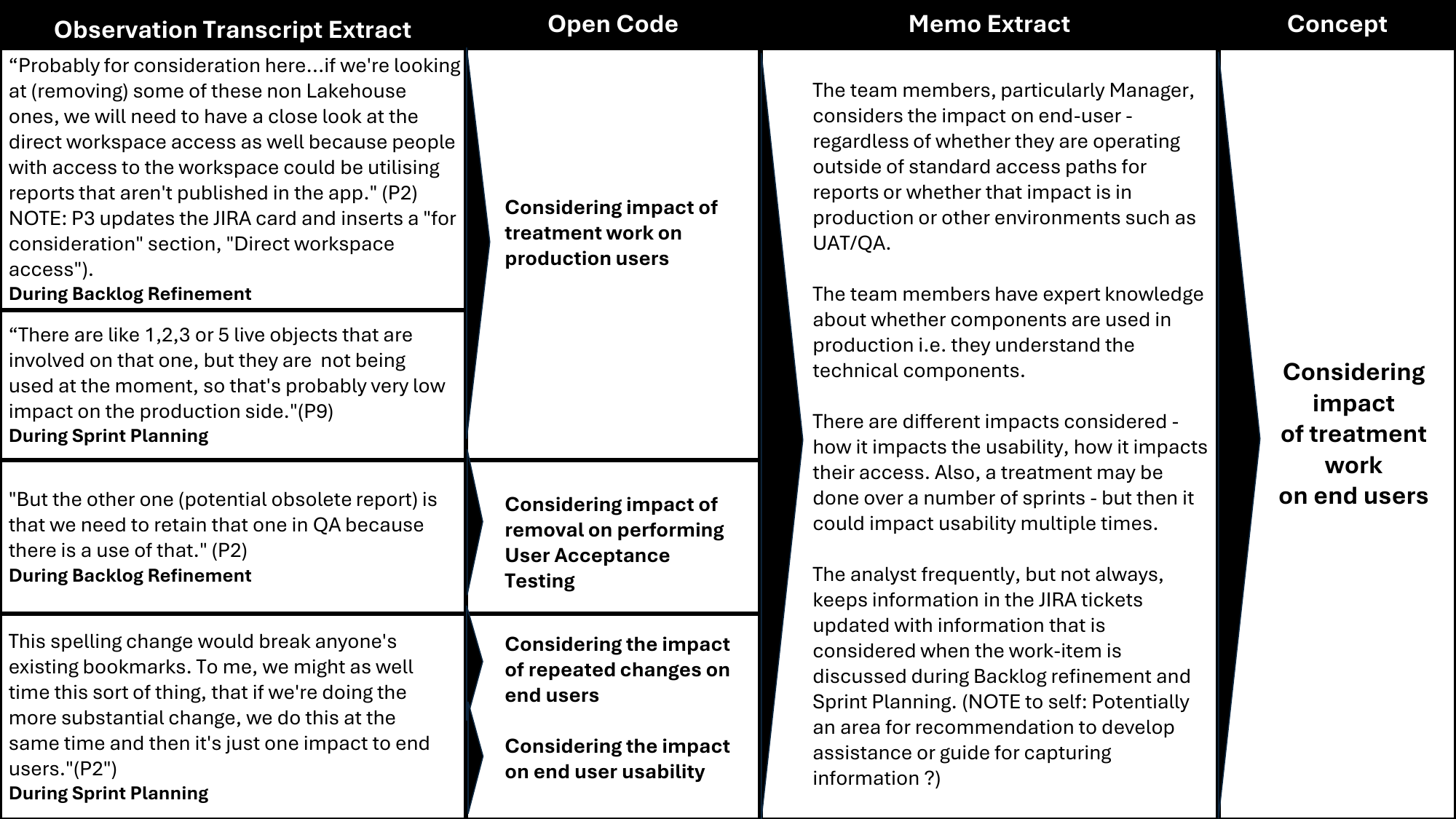}
\caption{Applying STGT for data analysis: Example of open coding and memoing to develop a concept}
\label{fig:CodeExample}
\end{figure*}

\subsection{Data Collection} \label {subsec5}
%\subsubsection{Collecting the Data}
 The first author was `onboarded' to the `Star Squad' by being invited to all ceremonies - Sprint Planning, Backlog review, Daily Standup (and associated After Parties), Sprint Review and Sprint Retro. She was also included in the team's Slack \footnote{https://slack.com} channel to be part of the day-to-day team collaboration messages.
 %The initial data collection plan included an initial Workshop to capture context information. However, as the team had just recently been formed by merging 3 partial teams in an organisational restructure, the `Star Squad' held a Team Kick-off workshop, facilitated by the Iteration Manager, and the first author proceeded to observe this meeting instead.\par
\textcolor{black}{Per the case study protocol, the first author had planned to conduct an initial workshop with the team to understand their roles and responsibilities and gather project and organisational context information. However, during the onboarding, the Iteration Manager explained that the team had just recently been restructured and formed by merging 3 partial teams in an organisational restructure and would be holding a Team Kick-off workshop, facilitated by the Iteration Manager, which would cover the contextual topics. This workshop was held on the third day of the observation study.}

\par \textcolor{black}{The first author agreed with the Iteration Manager that she could attend the Star Squad Kick-off workshop as an observer and then assess if a separate project context workshop was still required.  The team Kick-off workshop was conducted as an observation and covered substantial context information including the team structure, roles, relationship to other teams, organisational structure and a description of upcoming work.}
 
\par \textcolor{black}{After attending the team Kick-off, the first author determined that a separate workshop with the whole team was not required, and agreed with the Iteration Manager that she could cover off any remaining questions during clarification discussions.} 

\par \textcolor{black}{The project and organisational context information were reviewed and validated during the member checking process described in Section \ref{membercheck}.}

\par The team operated in hybrid mode and one or more team members were off-site for each meeting. Meetings were conducted as Zoom sessions. The first author `attended' the Zoom meeting with her `Zoom background' slide. Once participants provided consent, the Iteration Manager delegated control to manage recording to the first author (except for Sprint Review sessions). In addition to the scheduled ceremonies, the first author also attended additional Deep Dive sessions. These were selected based on discussions during daily stand-ups and covered a selection of analysis review discussions and release pipeline discussions. She made observational field notes during each observation, including information about the attendees at the session and any questions or issues arising out of discussions. She used these notes to formulate clarifying questions to be addressed with participants \textcolor{black}{one-on-one} in subsequent Clarification Sessions, which were also recorded. \textcolor{black}{In total, 15 such sessions were held over the 6 week period}.  The Clarification Sessions also provided an opportunity to capture education discipline and career background information from each participant. The observations commenced mid-sprint, with a Backlog Refinement session on January 30, 2024, and concluded with a Sprint Retrospective on March 20, 2024. The final reflective workshop/\textcolor{black}{focus group} was conducted on 22nd May 2024. We observed %93
\textcolor{black}{78} sessions with a total duration of 
%56
\textcolor{black}{49.5} hrs. A summary of the types, number of \textcolor{black}{observation} sessions and total duration of the observations is provided in Table 1.

\begin{figure*} [h]
\includegraphics[width=0.9\textwidth, center]{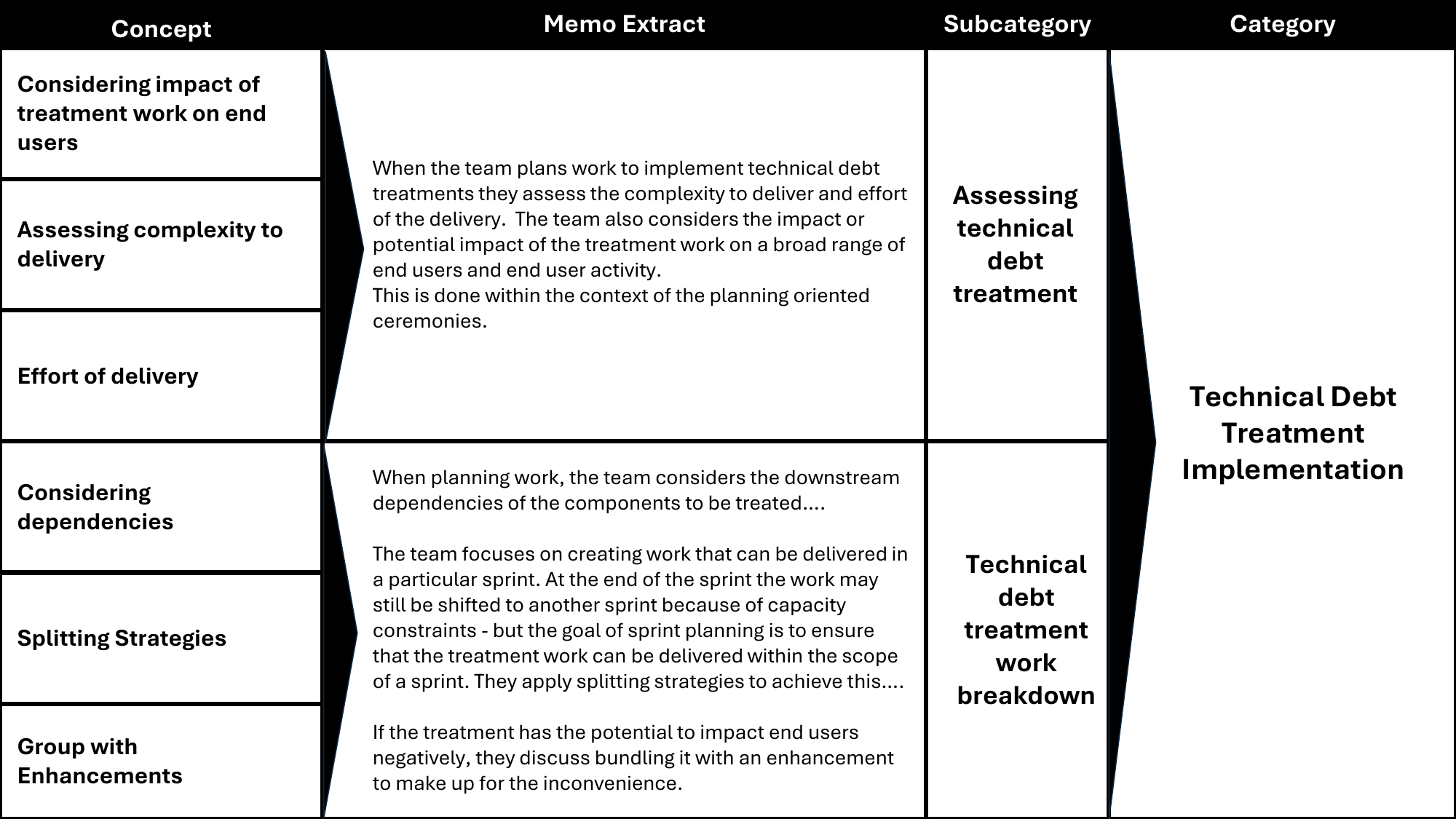}
\caption{Applying STGT for data analysis: Examples of categorising concepts into subcategories and category}
\label{fig:CategoryExample2}
\end{figure*}

\subsection{Data Analysis using Socio-Technical Grounded Theory} \label {subsec6}

Socio-Technical Grounded Theory (STGT) \citep{Hoda2022, Hoda24} is an inductive, qualitative research method with guidelines that adapts traditional Grounded Theory methods from the sociology domain \citep{GlaserBarney2017Tdog, CorbinJulietM.2008Boqr, CharmazKathy2014Cgt} to the software engineering domain. 
\textcolor{black}{We specificially selected} STGT \textcolor{black}{\textit{for data analysis} because we wanted to explore an underexplored phenomena\textcolor{black}{, i.e.,} the work practices of a multidisciplinary DI team that develops DI solutions. STGT} is specifically suited to \textcolor{black}{exploration of such} socio-technical phenomena. STGT expects researchers to have a socio-technical focus and ability to understand the language and general technical domain. Our research team are experts in software engineering and have skills and interest in human-centric research. Further, the first author is an experienced industry project practitioner. The researchers had a good grasp of the observed team's technologies and domain and were able to understand the domain language, technology and processes that the participants referred to.\par
STGT caters for the analysis of data from a variety of sources such as observations, images and logs. It provides an iterative approach to collect, analyse and structure data into concepts, categories and theories. STGT starts with a broad topic and narrows analysis down to key phenomenon (in case of exploratory studies with no specific focus) or key categories. STGT has two stages: a \textit{basic stage} for basic data collection and analysis, and an \textit{advanced stage} for mature theory development. It allows for limited application of the basic stage such as to analyse data collected using various methods. For this study, we used a limited application of STGT\textcolor{black}{, i.e.,} STGT \textit{for data analysis} to analyse the data collected within a case study construct \citep[Chapter 10]{Hoda24}. During data analysis it became apparent that managing \textcolor{black}{TD} in a multidisciplinary DI software team was one of the key phenomenon of interest. Concepts and categories conceptualising the types and properties of \textcolor{black}{TD}, types of treatments, and work planning to treat \textcolor{black}{TD} emerged early in the analysis. We continued to focus our analysis on this phenomena by aiming to answer the research questions outlined in the motivation section.  \par

The STGT method does not prescribe a \textcolor{black}{particular} research paradigm such as positivism. \textcolor{black}{Instead}, it expects the researchers to \textcolor{black}{select and} articulate their perspective \textcolor{black}{and use their expertise to interpret the data \citep[Chapter 5]{Hoda24}}. The first author, who conducted the data collection, clarifying interviews and analysis, has extensive industry project delivery experience. We wish to acknowledge the adoption of a subjective, context specific, constructivist research paradigm \textcolor{black}{for} constructing questions, interpretation of answers, formulation of concepts, categories and their interrelationships.\par

\subsubsection{Preparing and Filtering the Data}
Data preparation is ``the process of converting the raw data into formats (typically text-based) where qualitative analysis can be performed efficiently" \citep [Chapter 9] {Hoda24}. Data preparation for analysis involved converting each Zoom recording into an annotated, anonymised transcript. We used Otter.ai\footnote{https://otterai.com} to transcribe the Zoom recording, converting it into text. We anonymised the transcript by replacing team member names, and references to named stakeholders and organisations, and we annotated the transcript with key references to the video, including the Jira\footnote{https://www.atlassian.com/software/jira} ticket that the discussion referred to, or descriptions about how participants interacted with tools as part of a discussion or demonstration. The aim was to ensure that key visual information from the observation was connected to the transcript to provide context during the analysis.\par

Per our agreement with the organisation, we also anonymised the video recordings. We downloaded the Zoom recordings to Adobe Premiere Pro\footnote{https://www.adobe.com/au/creativecloud.html}, blurred identifying and commercially sensitive information and re-encoded the blurred video. We deleted the original. From the anonymised videos, we extracted relevant screenshots of Jira tickets (work items), Miro\footnote{https://miro.com/} boards, and Documents that had been shared and discussed during ceremonies per our case study design to enrich the analysis. We also extracted team Slack messages, ensuring that we captured each message and associated replies, at the end of each week into a spreadsheet. Per our agreement with the organisation, the anonymised videos and transcripts were made available to the participants (only the participants involved in the respective sessions), for a 2 week review period to provide participants the opportunity to raise any concerns, specifically regarding commercial sensitivity. Two participants provided feedback, thanking us for the opportunity to review.\par

One issue to highlight about data collection and anonymisation was the manual effort required. Whilst we used an automated transcription service, the associated manual effort was much higher than we anticipated and involved numerous corrections in addition to anonymisation. The meetings contained technical terminology, multiple speakers and a number of participants were not native English speakers. Every 1 hour of meeting required 2 hours of additional manual effort to correct and anonymise. Blurring of videos was also manually intensive, but necessary due to the privacy and commercial sensitivity of the material. Each hour of video required at least 1 - 2 hours of effort to anonymise, followed by encoding.

\par We imported the anonymised transcriptions and \textcolor{black}{Slack messages} into NVivo for \textcolor{black}{data preparation and filtering}.  Data filtering is “the process of identifying the key information, contextual information, and noise in the raw data” \citep [Chapter 9] {Hoda24}. To streamline later analysis of the data we decided that grouping by “work” was relevant information and so we elected to \textcolor{black}{connect each exchange or set of exchanges to the related Jira ticket where relevant.  Each Jira ticket was represented as a ‘case’ in NVivo.}

\par To achieve a ‘grouping’ by work, the first author coded those parts of a transcript file that were relevant to Jira tickets to the relevant NVivo ‘case’ record representing the Jira ticket. We analysed observation videos and took screenshots of the Jira ticket from the anonymised video along with additional screenshots of document contents, \textcolor{black}{Miro} boards, diagrams or examples of how team members interacted with their development tools. We assigned  screenshots to the relevant NVivo case, enabling a grouping by work.  We also annotated each transcript with references to the relevant screenshots e.g. P2 referenced the Miro board, P3 referenced Jira board).

\par To aid in familiarisation, the first author reviewed each observation transcript and made ‘annotations’ in \textcolor{black}{NVivo}. This involved highlighting segments of text, usually the dialog component of a particular team member or a dialog exchange between 2 or more team members about the same topic. The annotation was then written for the purpose of sense making\textcolor{black}{, i.e.,} to summarise and articulate the topic, the nature of the exchange and team members involved in the exchange - for example P4 provided data examples and requested clarification and P2 provided domain examples in response. The first author used the annotations to review and internalise material. The grouping by work enabled a sequential perspective of a piece of work, which had been somewhat difficult at times during the actual observations because of the speed of the actual exchanges. Whilst annotating and internalising, the first author commenced memo writing to capture reflections about key information, including organisational context information, data work practices and TD work.

\par\textcolor{black}{To assess the instruments for reliability, during the familiarisation process, the first author met with the co-authors to discuss the approach in general and then met separately with the 3rd author for a more detailed review of the annotations and data connections in NVivo. The 3rd author found the process very comprehensive and detailed. The first author sought further guidance from the 2nd author who advised on a structured approach to data preparation and filtering \citep [Chapter 9] {Hoda24}. The first and third authors assessed the approach taken so far against the structured approach and determined that the grouping and data connections met the preparation requirements but filtering criteria should be enhanced to ensure that only relevant segments of the observation were coded and included in the analysis. Filtering criteria were then set to exclude exchanges that did not pertain to work that commenced during the observation project (e.g. Jira tickets that were in the closing stages at commencement of the observation study) and detailed domain specific discussions. The reason for excluding this work is that the observations did not cover any team discussions about the nature or scope of the work and the work was not further referenced during the observation study), and non development work related team member exchanges such as detailed domain discussions, social activities, personal and organisational policy matters.}

% and expert on socio technical grounded theory method,

\subsubsection{Data Analysis}
To analyse the data we open-coded the \textcolor{black}{TD} related data \textcolor{black}{consisting of} transcripts, screenshots and \textcolor{black}{Slack messages}. We used the coding feature in NVivo \citep{Hutchison2010, Soliman2004}. During open coding, we filtered out exchanges that did not pertain to work that commenced during the observation project including discussions about Jira tickets that were being completed in the early days of the observations, detailed discussions about the domain, and non work related team member exchanges such as social activities and personal matters.
\par
 Our analysis quickly identified that the observations of team ceremonies provided a rich dataset to surface \textcolor{black}{TD} characteristics and activities relating to how the \textcolor{black}{TD} was identified, and how solutions were selected and planned for implementation. We therefore established the key category of `Managing Technical Debt in a Multidisciplinary Data-Intensive Software Team' to provide a focal point for our analysis.\par
 We demonstrate the application of STGT's basic data analysis with a worked example in Figures \ref{fig:CodeExample} and \ref{fig:CategoryExample2}. \textcolor{black}{The first author performed open coding of data by reviewing each transcript segment (consisting of one or more sentences from a participant and sometimes an exchange by participants), Slack messages in a thread, and extracted video screenshots. During open coding, the first author sought to understand what was happening, what actions were taking place and what tools were being used. To build this understanding, the first author reviewed the data connections (e.g. screenshots and in the case of Jira tickets, prior discussions and context about the work item). An example of open codes assigned is shown in the left most column of Figure 1. During the open coding process, memoing continued to capture reflections relating the codes to each other into concepts.}

A worked example of how open codes are assigned to build concepts through basic memos is shown in column 2 of Figure \ref{fig:CodeExample}. More conceptual memos \textcolor{black}{were} then developed to structure codes into subcategories and categories, as demonstrated in \textcolor{black}{column two} of Figure \ref{fig:CategoryExample2}. Diagramming was then used to visualise and further structure the categories and concepts. The first author regularly met with the others to discuss the analysis approach and review coding examples and progress. \textcolor{black}{During review meetings, the authors asked clarifying questions and agreed with the approach taken.  If there was disagreement, the authors sought consensus through discussion. For example, on one particular occasion where the authors reviewed the category and concepts relating to treatment of data and \textcolor{black}{TD}, there was discussion about the category definition at the time being overloaded. Each author stated their perspective and reasoning and the authors reached a consensus that having 2 categories, one to focus on treatments for technical data debt, and one to focus on treatment implementations may be more appropriate. The first author updated the analysis and reviewed the results with the co-authors and continued with that analysis.}
\par
The results of the analysis in terms of identified concepts and categories are \textcolor{black}{visualised in} Figure \ref{fig:ConceptVisualisationd} which shows the concepts, subcategories and categories identified within the overall case context. We also identify a concept called \textit{evolution} on the boundaries between the case context and case to indicate ongoing and evolving impacts from the context to the unit of analysis. The case context and detailed conceptual findings (including the concept of \textit{evolution}) is explained in Section\textcolor{black}{ 4}.
\par \textcolor{black}{Note: our findings are not documented in terms of advanced financial technical debt language. This reflects the focus of the analysis, which was the team under observation and reflects the nature of technical debt language used by the participants. Our analysis findings are mapped to technical debt terminology and situated in existing TD literature as part of our discussions in Section 6.2.}

\textcolor{black}{\subsubsection{Member checking and commercial sensitivity check}} \label{membercheck}
\textcolor{black}{At completion of the analysis and after drafting the research paper, the first author performed a member check process and offered participants the opportunity to review the quotes that were being considered for publication in the journal paper, and a presentation of the observation analysis results. Five of the team members took up the offer for a presentation and discussion. The member check discussions were not treated as observations, were not recorded and apart from corrections, did not form part of the analysis.} 
\par \textcolor{black}{The team members found the case context representative and the TD concepts and category findings insightful and representative of the work they did. Two team members expressed that the findings and terminology could help them improve their practices, as this could be useful for guiding and structuring team discussions about \textcolor{black}{TD} and practice improvement. As these member check discussions were not part of the analysis, we do not make any claims about the usefulness of the framework. Two team members provided corrections to case context information -  about the overall architecture and Scrum terminology. We updated the case context with the corrected information. After completing the checks with individual team members the first author submitted a copy of the draft journal submission to the team manager, to conduct a final commercial sensitivity check and received confirmation that the paper would not pose a risk from a commercial sensitivity perspective.}\\ 

\section{Findings}
\subsection{Case Study Context}
%In this section we present the socio-technical context of the case.
\subsubsection{Organisation Structures}\label{OrgStructure}
\begin{figure*} [h!]
\includegraphics[width=0.6\textwidth, center]{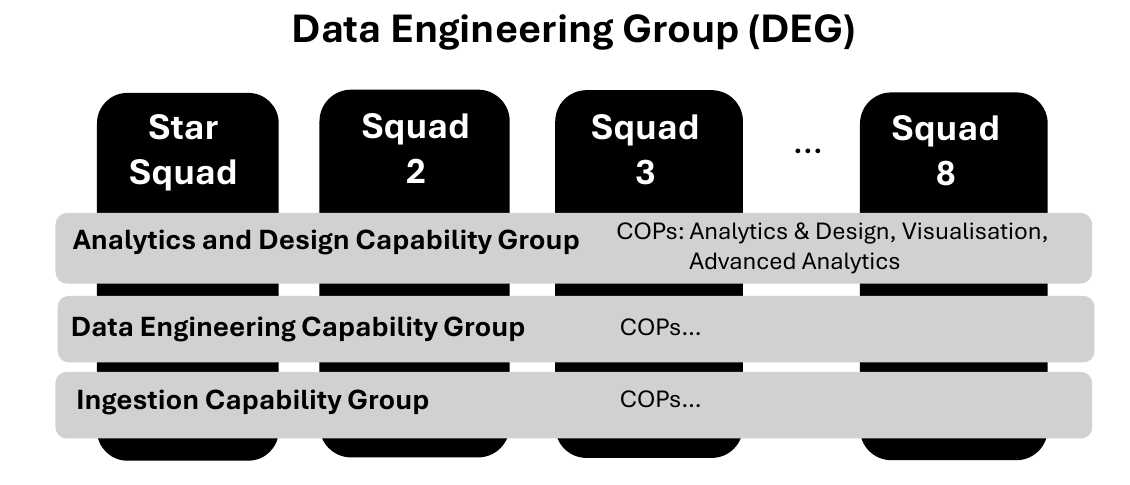}
\caption{Context: Star Squad Organisation Context}
\label{fig:Organisation}
\end{figure*}

The observed `Star Squad' was responsible for the delivery and support of the Enterprise Reporting system. The observed team belonged to the Data Engineering Group (DEG) in the organisation's broader IT Group. The observed team was one of 8 teams in DEG, and there were 12 team members out of a total of 73 across the 8 teams. The observed team members and roles are outlined in Table \ref{tab:teamroles}. Team members were allocated to capability groups and community of practices (COPs) aligned with their roles and interests. For example, analysts in the observed team members belonged to one of the Analytics and Design, Data Engineering, or Ingestion capability groups. Within the Analytics and Design, there were three COPs - Data Analytics and Design, Visualisation, and Advanced Analytics. The operation of the other teams, COPs, and Capability groups were not included in the observed scope, but are included for context.  See Figure \ref{fig:Organisation} for an overview of the organisation structure.

\begin{figure*} [h]
\includegraphics[width=0.6\textwidth, center]{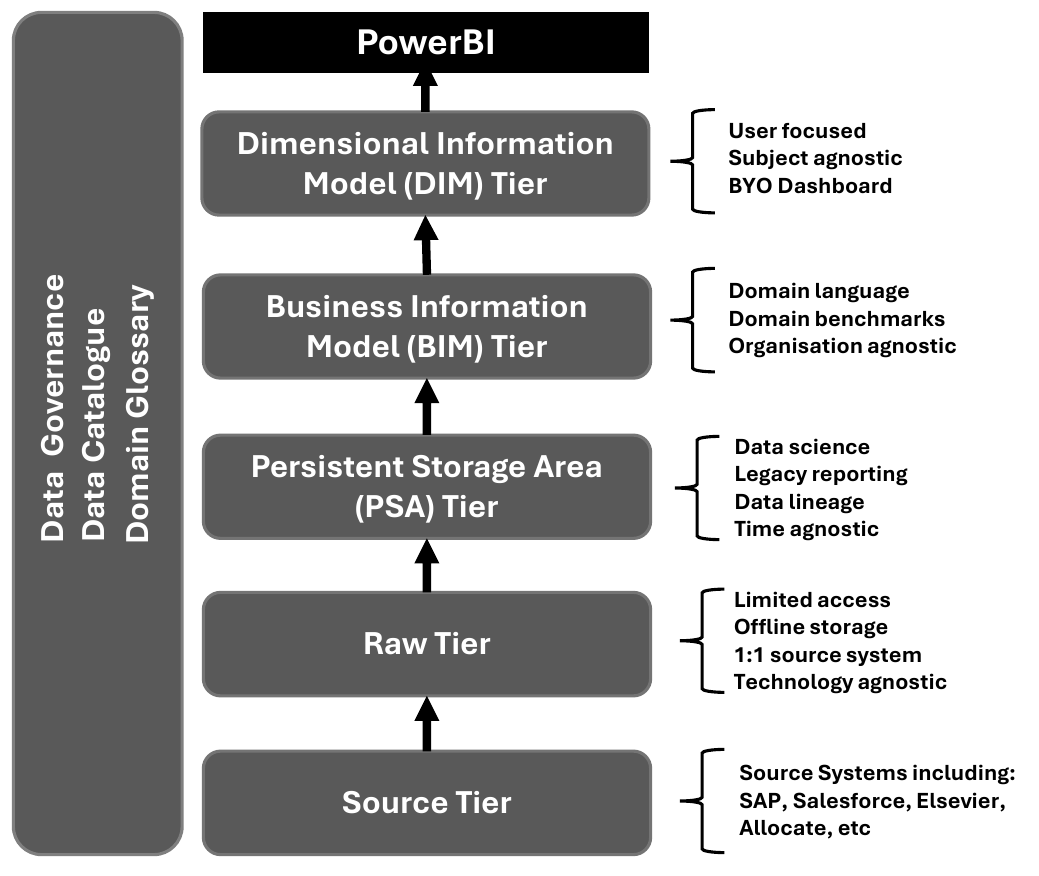}
\caption{Context: (Anonymised) Overview of Organisation Datawarehouse Architecture Tiers}
\label{fig:Architecture}
\end{figure*}

\subsubsection{Product and Stakeholders}
The enterprise reporting end product consisted of a number of PowerBI\footnote{https://www.microsoft.com/en-us/power-platform/products/power-bi} dashboards deployed through workspaces, as well as access to the underlying datasets (available to power users) and personalisations (available to all PowerBI users. The end user community was made up of data analysts as well as senior management and executive stakeholders. Data analysts who had different levels of analytical skills had access to the dashboards and datasets, whereas management level stakeholders generally had access to dashboards and perhaps personalisation features. End users, mainly data analyst end users, were connected to the observed team through COPs and also engaged as ‘domain experts’ in specialised groups for consultation and user acceptance testing.  The team also regularly engaged with a data governor. 

\subsubsection{Technology, Architecture and Tools}
The observed team works in a Business Intelligence Enterprise Reporting Datawarehouse environment \citep{kimball2008DWL}. DEG is in the process of implementing the Databricks\footnote{https://www.databricks.com/} Lakehouse Platform. This platform enables a tiered Data Warehouse Architecture as shows in Figure \ref{fig:Architecture}, which connects multiple source systems through the base tiers responsible for processing the raw data (Raw Tier) and separating out personally identifiable information from other information into the Persistent Storage Area (PSA Tier). These lower level tiers are outside the scope of our study as they are developed and maintained by other DEG teams. The Business Information Model (BIM) Tier provides models set up in business language. It feeds central dashboards, whereas the Dimensional Information Model (DIM) provides the final Tier to support end user accessible dashboards and datasets accessible through Microsoft PowerBI models and visualisations. The organisation has established Data Governance processes as well as Data Catalogues and Business Glossary. The details and operation of these processes and structures were not included in the case study observation but are noted here for context.\par

The BIM and DIM layers in the Lakehouse Architecture had been developed incrementally over a period of 18 months. The development teams had recently been restructured into the structure of 8 teams noted in Section \ref{OrgStructure}, including the `Star Squad', which was the subject of observation.\par

We observed the `Star Squad' members use several tools, including Databricks Catalogues and Notebooks for SQL development (BIM and DIM), Microsoft PowerBI for data visualisation and dashboard development, and Azure DevOps Pipelines\footnote{https://azure.microsoft.com/en-us/products/devops/pipelines/} for BIM and DIM workflow processing and release deployments. PowerBI Deployment pipelines are used for PowerBI deployments. Importantly, the Azure DevOps Pipelines and PowerBI pipelines were not yet integrated due to technical limitations, which contributed to challenges elaborated in Section 4.2 below. The team used Lucidchart\footnote{https://www.lucidchart.com/pages/} for data modeling and Atlassian Confluence\footnote{https://www.atlassian.com/software/confluence} to document BIM designs.
The team used Atlassian JIRA\footnote{https://www.atlassian.com/software/jira} to track, manage, and report work in sprints. Outside of team ceremonies, day-to-day team collaboration was done through Slack\footnote{https://slack.com/intl/en-au/}, and team ceremonies were always in hybrid mode, conducted over Zoom\footnote{https://zoom.us/}. At times, meetings were
facilitated through Miro and associated templates\footnote{https://miro.com/}. The team also had access to Google Workspace tools\footnote{https://workspace.google.com/intl/en-au/}.

\subsubsection{Team Members}
The Star Squad was a multidisciplinary team, consisting of 8 data analysts, a product manager,  iteration manager, business analyst, and quality assurance expert, as shown in Table \ref{tab:teamroles}. The team members' education discipline backgrounds are shown in Table \ref{tab:Education Discipline}, and their career backgrounds are summarised in Table \ref{tab:Career Background}.  Some team members had education backgrounds in more than one discipline. The discipline backgrounds span a wide range of domains, including  non technology related domains, general information technology, engineering, data science, computer science\textcolor{black}{,} data analytics, and management. The Product Manager, in particular, had established deep domain knowledge in the organisations domain, with most of their professional \textcolor{black}{experience} gained working in similar organisations. They also had a deep knowledge of the organisation's legacy application environment. Analyst team members were largely aligned to tasks relating to visualisations,  or backend data analysis and query development.  The team had a balance of very deep level of professional expertise and junior team members (See Table \ref{tab:Professional Experience}). Note, the information in team members related Tables \ref{tab:teamroles}, \ref{tab:Education Discipline}, \ref{tab:Professional Experience}, and \ref{tab:Career Background}, has been generalised and aggregated to preserve team member privacy.

\begin{table}[h]%% placement specifier
%  \vspace{-2.5mm}
  \footnotesize
% \renewcommand{\arraystretch}{1.5}
 %\resizebox{18cm}{!}{
\caption{\textcolor{black}{\\Participant Team Member Roles}}
\vspace{-0.3cm} 
  \begin{tabular}{p{1.3cm} p{4cm}}
  \hline
\textbf{ID} & \textbf{Role (Focus Area)}\\ %% A tabular row ends with \\
 \hline % \hline
 P1& Manager \\
% \hline
 P2& Manager \\
% \hline
 P3& Analyst \\ 
% \hline
 P4& Data Analyst - Backend\\ 
% \hline
 P5& Data Analyst - Platform \\ 
% \hline
 P6& Data Analyst - Visualisation\\ 
% \hline
 P7& Data Analyst - Visualisation \\ 
% \hline
 P8& Data Analyst - Backend\\ 
% \hline
 P9& Data Analyst - Backend\\ 
% \hline
 P10& Data Analyst \\ 
% \hline
 P11& Analyst \\
% \hline
 P12& Data Analyst - Visualisation\\ 
 \hline
\end{tabular}
% \caption{Participant Team Member Roles}
  \label{tab:teamroles}
\end{table}

\begin{table}[h]%% placement specifier
%  \vspace{-2.5mm}
  \footnotesize
  \caption{\textcolor{black}{\\Aggregated Participant Team Member Education Disciplines}}
\vspace{-0.3cm} 
 %\renewcommand{\arraystretch}{1.5}
 %\resizebox{18cm}{!}{
  \begin{tabular}{p{4cm}  c }
  \hline
\textbf{Education Discipline} & \textbf{No. of Team Members} \\ %% A tabular row ends with \\
 \hline %\hline
 Domain & 2 \\
% \hline
 Information Technology & 2 \\ 
% \hline
 Data Analytics & 2 \\ 
% \hline
 Engineering& 3 \\
% \hline
 Computer Science& 2\\
% \hline
 Data Science& 2 \\
% \hline
 Training & 1\\
% \hline
 Business Management & 1\\
\hline
\end{tabular}
% \caption{Aggregated Participant Team Member Education Disciplines}
  \label{tab:Education Discipline}
\end{table}

\begin{table}[h]%% placement specifier
%  \vspace{-2.5mm}
  \footnotesize
   \caption{\textcolor{black}{\\Aggregated Participant Team Member Years Professional Experience}}
   \vspace{-0.3cm} 
% \renewcommand{\arraystretch}{1.5}
% \resizebox{18cm}{!}{
  \begin{tabular}{ c  c }
  \hline
\textbf{Years of Experience} & \textbf{No. of Team Members}\\ %% A tabular row ends with \\
 \hline %\hline
 $<$ 2 & 2\\
% \hline
 2 - 4 & 0\\
% \hline
 5-10 & 5\\
% \hline
 $>$10 & 5\\
 \hline
\end{tabular}
% \caption{Aggregated Participant Team Member Years Professional Experience}
  \label{tab:Professional Experience}
\end{table}

\begin{table}[h]%% placement specifier
%  \vspace{-2.5mm}
  \footnotesize
  \caption{\textcolor{black}{\\Aggregated Participant Team Member Career Backgrounds}}
   \vspace{-0.3cm} 
% \renewcommand{\arraystretch}{1.5}
% \resizebox{18cm}{!}{
  \begin{tabular}{p{4cm}  c }
  \hline
 \textbf{Career Backgrounds} & \textbf{No. of Team Members}\\ %% A tabular row ends with \\
 \hline %\hline
Business Analysis & 2\\
%\hline
Data Engineering & 5\\
%\hline
Data Modeling & 2\\
%\hline
Data Reporting and Visualisation  & 3\\
% \hline
IT Training and Support & 2\\
%\hline
Product/Project Management & 3\\
%\hline
Quality Assurance & 1\\
\hline
\end{tabular}
% \caption{Aggregated Participant Team Member Career Backgrounds}
  \label{tab:Career Background}
\end{table}

\begin{figure*} [h]
\includegraphics[width=1.0\textwidth, center]{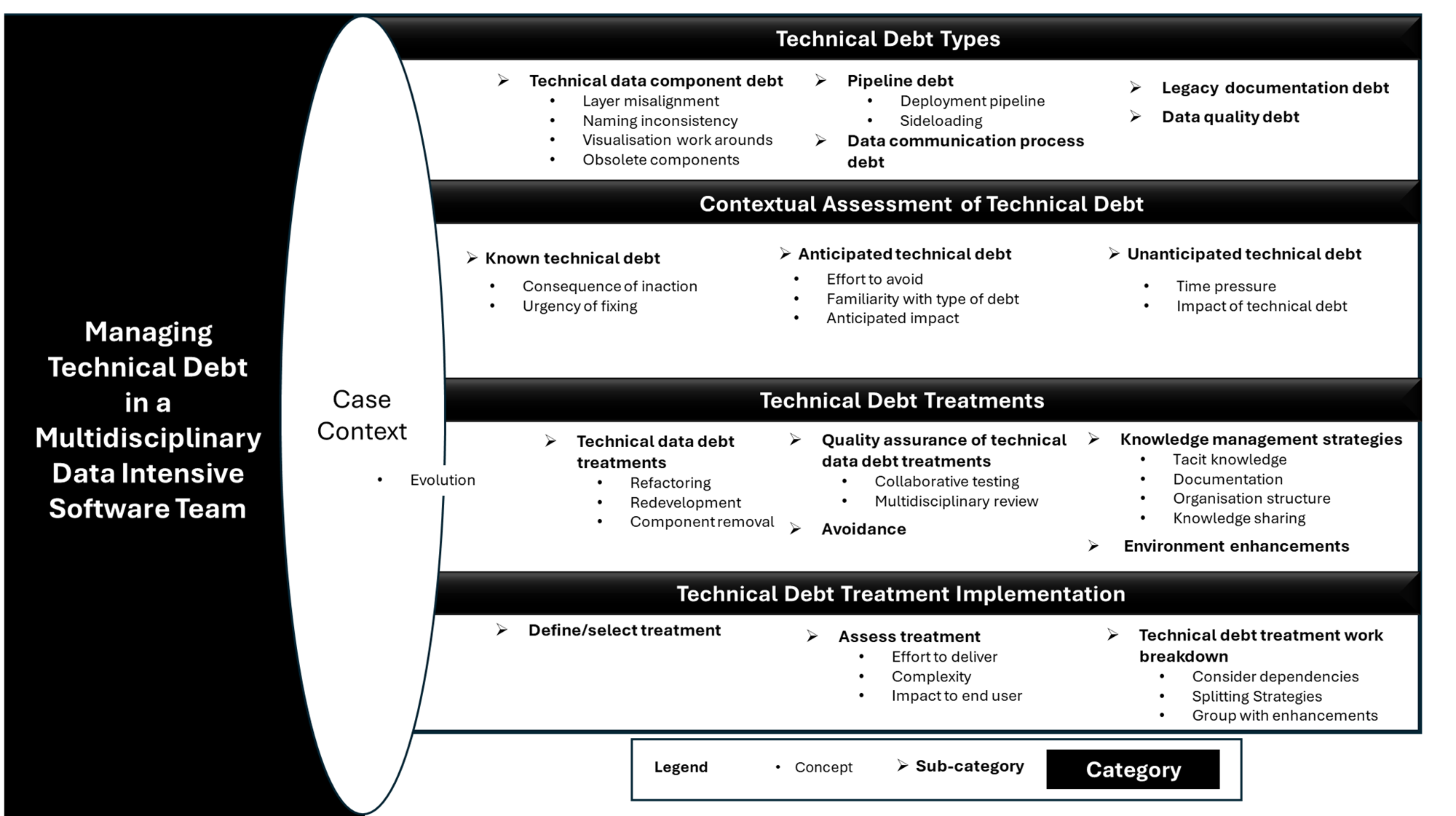}
\caption{\textcolor{black}{Visualisation} of Managing technical debt in a data-intensive software team} 
\label{fig:ConceptVisualisationd}
\end{figure*}

\subsubsection{Ways of Working}
\textbf{Ceremonies:}
The 12 member team used %agile
scrum practices to manage their work. They held the following ceremonies (with time ranges) fortnightly: a) \textit{backlog refinement} (1hr - 1.5hrs) to scope out, estimate and prioritise work items, b) \textit{sprint planning} (3hrs) to close out the existing sprint or carry over work from the prior sprint and finalise (scope, estimate and prioritise) the items for the commencing sprint, c) \textit{sprint review} (30min-1hr) to review the previous sprint and showcase deliverables within the team and to external stakeholders, and d) \textit{sprint retrospective} (1hr) to review sprint performance and conduct reflection on how to improve.  Every day, except on sprint-planning days, the team held a \textit{stand-up} (10min - 30min) ceremony to discuss progress and, if required, raised the need for an \textit{after-party} (10min - 50min) ceremony to discuss issues, next steps, or conduct reviews with each other and/or the product manager. All ceremonies were  attended by all team members, except after-parties, which were held on most days and only attended by those team members involved in the discussion.\par
\textbf{Observed Data-Intensive Work:}
For the purpose of contextualising, we characterise the observed DI work performed by team members as: a) \textit{data development}, b) \textit{data visualisation development}, c) \textit{quality assurance}, d) \textit{data deployment}, and e) \textit{Visualisation deployment} work to deliver the enterprise business intelligence product. For the purpose of this paper, we define \textit{data development} work as including the SQL and related coding work including the design, creation and updating of Business Information Models (BIM) and Dimensional Data Model (DIM) data tiers in the architecture. We define \textit{data visualisation development} work as the data modeling and visualisation development in PowerBI. Quality assurance work activities included testing and review activities associated with \textit{data development}, \textit{data visualisation development} and their integration. \textit{Data deployment} work includes the maintenance and continuous updating of workflows and pipelines to accommodate changes in the BIM and DIM data into the Development (DEV), Quality Assurance (QA) and Production (PROD) environments. \textit{Visualisation deployment} includes the deployment of PowerBI dashboards and refreshing of data into the data visualisations in the DEV, QA and PROD environments.\par Team members were observed to perform or discuss the performances of DI work shown in Table \ref{tab:DI work}.\\
\begin{table*}[h]%% placement specifier
   \footnotesize
    \captionsetup{justification=raggedright, singlelinecheck=false}
   \caption{\textcolor{black}{\\DI Work performance observed or discussed by team members}}
   \vspace{-0.3cm} 
 %  \centering
  
% \renewcommand{\arraystretch}{1.5}
%  \resizebox{18cm}{!}{
  
  \begin{tabular}{c c c c c c }

%\begin{tabularx}{1.0\textwidth} {c c c c c c }
  \hline
\textbf{  Participant   } & \textbf{   Data Development   } & \textbf{Data Visualisation} & \textbf{Quality Assurance} & \textbf{Data Deployment} &\textbf{Visualisation Deployment}\\
\hline
\textcolor{black}{P1} & &  & \textcolor{black}{$\medbullet$ }&  &  \\
 \hline
  P2 & $\medbullet$ &  & $\medbullet$  &  &  \\
 \hline
 P3 &  &  &  &  &   \\
 \hline
 P4 & $\medbullet$ &   & $\medbullet$ & $\medbullet$ &  \\
 \hline
 P5 & $\medbullet$ &   & $\medbullet$ &   &  \\
 \hline
 P6 &   & $\medbullet$ & $\medbullet$ &   & $\medbullet$ \\
 \hline
 P7 & $\medbullet$ & $\medbullet$ & $\medbullet$ & $\medbullet$ & $\medbullet$\\
 \hline
 P8 & $\medbullet$ &  & $\medbullet$ & $\medbullet$ &   \\
 \hline
 P9 & $\medbullet$ &  & $\medbullet$ & $\medbullet$ &   \\
 \hline
 P10 & $\medbullet$ & $\medbullet$ & $\medbullet$ &  & $\medbullet$ \\
 \hline
 P11 &   &   & $\medbullet$ & $\medbullet$ & $\medbullet$ \\
 \hline
 P12 &   & $\medbullet$ & $\medbullet$ &  & $\medbullet$ \\
%% A tabular row ends with \\
 \hline 
 \end{tabular}
%\caption{DI Work performance observed or discussed by team members}
\label{tab:DI work}
\end{table*}

\textbf{Measured Work Types:}
DEG teams were measured through various metrics discussed at leadership meetings (outside the scope of the observation study). The metrics were collected through Jira work item types, supplemented by highlight descriptions provided by the product manager and iteration manager. For the context of the findings related to \textcolor{black}{TD}, relevant information includes that the measured  `Work Item Type' and \% range of work completed each month over the prior 4 months included \textit{Break Fix:} $<$5\%, \textit{continuous Improvement:} 8\% - 9\% , \textit{discovery incl. analysis: 0 - 9\%} and \textit{value work: 80\%- 90\% }. Further, the amount of \textit{continuous improvement} work was kept relatively stable, ranging between  8\% - 9.5\% per month, closely related to and balanced with value work. Continuous improvement included enhancement work items as well as \textcolor{black}{TD} work items. The team was not observed to have a formal register for maintaining \textcolor{black}{TD}.

\subsection{Managing Technical Debt in a Data-Intensive Software Team}
We conceptualised several categories and concepts around our key category of \textbf{Managing Technical Debt in a Data-Intensive Software Team}, which we present in this section. The category \textbf{Technical Debt Types} is made up of sub-categories that describe the \textbf{Technical Debt Types}. We identified 3 further categories which conceptualise activities relating to \textcolor{black}{TD} management, namely \textbf{Identify and Assess \textcolor{black}{TD}}, \textbf{\textcolor{black}{TD} Treatments} and \textbf{\textcolor{black}{TD} Treatment Work Breakdown}.\par

For each category, we present related sub-categories and concepts, some illustrated by sanitised quotes and conversational extracts. \textcolor{black}{Unless otherwise noted, quotes are conversational extracts from observations.} \textcolor{black}{We use \faComment \hspace{0cm} in conjunction with the participant identifier (Px, role) in the following subsections. Where more than one participant continues in the dialogue, we omit \faComment \hspace{0cm} before their identifier.} Figure \ref{fig:ConceptVisualisationd} provides a visualisation of this conceptualisation. We present a summary at the end of each category in answer \textcolor{black}{to} our research questions.

\subsubsection{Technical Debt Types}
We observed several references and discussions relating to \textcolor{black}{TD} that we separated into \textbf{Technical Debt Types}, namely \textcolor{black}{\textbf{technical data component debt}}, \textbf {pipeline debt}, \textcolor{black}{\textbf{data communication process debt}}, \textbf{data quality debt} and \textcolor{black}{\textbf{legacy documentation debt}} sub-categories.
\par
\textcolor{black}{\textbf{Technical data component debt}} encompasses debt relating to BIM, DIM and PowerBI \textit{layer alignment} where data is accessed directly from BIM structures (because the DIM layer has not been created), or the data layers still contain references to  legacy components. The \textcolor{black}{sub-category} also includes \textcolor{black}{the concept} \textit{naming inconsistency} covering inconsistent naming data fields or report naming conventions, and DIM naming issues. For example:

\faComment \hspace{0cm}  P2 (Manager): \textit{``to be able to do this, we would be referencing BIM tables...and there is a need to assess the Tech Debt that we would create, by doing this piece directly off the BIM."}\\
\par
We also include concepts such as \textit{obsolete components} and Power BI \textit{visualisation workarounds} within this sub-category. \textit{Visualisation workarounds} \textcolor{black}{include} the need to implement modifications within PowerBI to achieve desired visuals due to underlying data structure or layer issues. An important impact of implementing modifications within the PowerBI visualisation layer is that access restrictions may need to be placed on access to the underlying dataset, impacting dataset access or personalisation by end users. \textcolor{black}{The \textit{obsolete components} concept represent components that have been superseded by new components but had not been removed from the environment and they remain dormant in the environment.}
\par
\textcolor{black}{\textbf{Pipeline debt} represents immature technology solutions that could not be automated fully and have to be supplemented with manual processes.}  Within the sub-category of \textcolor{black}{\textbf{pipeline debt}}, we identified 2 main concepts - \textit{deployment pipeline} and \textit{sideloading}. \textit{Deployment pipeline} related discussions centred mainly around issues with the deployment pipelines and data refreshes for PowerBI, which were not integrated with the Databricks Azure DevOps deployment pipelines. Workflows to `build' data constructs ran on an automated schedule within the Azure DevOps environment, but PowerBI model refreshes were scheduled or executed ad-hoc through the PowerBI Service. Due to technical platform limitations, integration of these pipelines could not be automated, and running selective Azure Ops orchestrations without PowerBI refresh caused downstream data alignment issues that resulted in outdated `keys' joining the FACT and DIM tables in datasets, which in turn resulted in misaligned datasets and non-sensical data flowing through to PowerBI dashboard \textcolor{black}{visualisations}. The team had implemented a manual co-ordination process via the team Slack channel to notify team members prior to running Azure Workflows, and this would then alert team members developing visualisations that they would need to refresh: \par 
\faComment \hspace{0cm} P8 (DA-Backend): \say{\textit{I'm about to kick-off dev qa orchestration. Please let me know if you have any concerns. Thanks!\\...Orchestration completed.}} \textcolor{black}{(sourced from Slack Channel)}\\
\par
Over time, the approach to deployment pipelines for applications had changed, leaving different approaches for older applications.\par
\faComment \hspace{0cm} P2 (Manager): \textit{``Do we have deployment pipelines for those reports?"}
\vspace{-0.3cm}\\
\par
 P7 (DA-Visualisation): \textit{\say{Not consistent, that was something that we implemented when we first started.}}
 \vspace{-0.3cm}\\
\par
 P12 (DA-Visualisation): \textit{\say{In the release meeting (with the other teams) it was mentioned that when we release some of these reports, we don't have a pipeline.}}\\
\par
Whilst most data consumed by the visualisations was sourced through source systems, the Lakehouse architecture also supported \textit{sideloading} of data files created through external (manual processes) necessary to fill gaps in source systems. \textcolor{black}{We categorised} \textit{sideloading} \textcolor{black}{as} a form of \textbf{pipeline debt} because the associated pipeline was identified as not having support for roll-back or archiving. The reason for requiring \textit{sideloading} is that historically, the enterprise reporting team took on responsibility for managing some reference data on behalf of business stakeholders using Excel files, and those processes never progressed through digital transformation:
\par
%\faComment \hspace{0cm} \textit{``We've got that consolidated into a single Excel file, because that's how it was being managed in Legacy. We were using multiple sheets to manage the data and then individually export those out as a CSV."} (P2-PM)\\
\faComment \hspace{0cm} P8 (DA - Backend): \say{\textit{I had a look at reference sheet. You mentioned about getting back to the `previous version'. So, if we have to get back to a previous version, because it has multiple tabs, will it be the case that we have to revert back all of those references to the previous version.}}\\
%\textit{``Some of those are linked, there's business keys in there that join between those."} (P2-PM).\\
%\textit{``So even if one file is wrong, everything has to be rolled back. Do we manage this, or does the business update this?"} (P8-DE)\\
%\textit{``We manage it."}(P2-PM)\\ 
\par
As source systems are updated and their data is integrated to the Lakehouse, effort needs to be expended to assess whether the \textit{sideloading} can be replaced with direct access:
\par
\faComment \hspace{0cm} P2 (Manager): \textit {``So rather than us manually uploading that as a .csv file into our platform, like we used to do, we might be able to source this from $<$ source system $>$ instead."}\\
\par
The team had practices for incorporating documentation for analysis, design and testing. However, there was missing documentation for some non-standard, legacy applications and pipelines and this \textcolor{black}{was categoriesed as} \textbf{\textcolor{black}{legacy} documentation debt}.  \textcolor{black}{Whilst the team had documentation practices in place for new development, these gaps related to legacy components. This needed to be addressed when one of the team members that had knowledge about these components} announced they were leaving to take up a new job opportunity.  \textbf{\textcolor{black}{Data communication} process debt} related discussions were observed regarding \textcolor{black}{inconsistencies in} processes and features for managing communication to end user stakeholders about data related issues:\par
\faComment \hspace{0cm} P1 (Manager): \textit{``We've taken a few different approaches with some of the dashboards (for communicating with end users), there's different levels. There's removing access to the dashboard and replacing it with the Service Desk message. Then we've got that `news box' on the landing page (for one of the dashboards), and on the (other dashboard) we just put text directly on to that landing page to communicate things previously as well. So we don't have a real consistency with our production workspaces at this point."}
\vspace{-0.3cm}\\
\par
\textcolor{black}{\faComment \hspace{0cm} P2 (Manager): \textit{``We don't have a sort of defined process really with, you know, updating text or the role of our team versus the [Other team] around managing some of this."}}
\par
The team also discussed \textbf{data quality debt}, mainly in relation to historical \textcolor{black}{data from legacy systems that lacked accuracy. The data quality debt sub-category also conceptualises discussions about missing data, duplicated records, and data alignment issues (that make it difficult to align data across different systems and tables).% The team established a practice to filter out pre 2005 data from the reporting layers.
} 

\begin{tcolorbox}[enhanced,attach boxed title to top left={yshift=-3mm,yshifttext=-1mm},
  colback=black!5!white,colframe=black!75!black,colbacktitle=black!80!black,
  title=Summary,fonttitle=\bfseries,
  boxed title style={size=small,colframe=black!50!black} ]
   \textbf{\textit{RQ1: What does a multidisciplinary, data-intensive system engineering team discuss about \textcolor{black}{TD}?}}\\
%  \hspace{0.5}
  We identified discussions about \textbf{technical \textcolor{black}{data component} debt}, \textbf{data quality debt}\textcolor{black}{, }\textbf{pipeline debt}, \textbf{\textcolor{black}{data communication} process debt} and \textbf{\textcolor{black}{legacy} documentation debt} sub-categories of \textcolor{black}{TD}. \textcolor{black}{The team members articulated that} \textbf{technical \textcolor{black}{data component} debt}, \textbf{\textcolor{black}{data communication} process debt}, and \textbf{\textcolor{black}{legacy} documentation debt} resulted from shortcuts taken during development. \textcolor{black}{\textbf{Pipeline debt}} \textcolor{black}{was experienced due to shortcomings in existing technology components.} \textcolor{black}{\textbf{Data quality debt} was experienced due to the nature of data, specifically legacy data.} \textbf{Pipeline debt} in particular was discussed in terms of, and managed through, manual processes.
\end{tcolorbox}

\subsubsection{Identify and Assess TD}
As part of their work delivery and planning, we observed that the team regularly identified, discussed and assessed \textcolor{black}{TD in their ceremonies, but the aspects of the TD that they considered and assessed were different based on the context}. We established three sub-categories based on the level of awareness \textcolor{black}{and} anticipation of \textcolor{black}{TD - \textbf{known TD}, \textbf{anticipated TD} and \textbf{unanticipated TD.} \textbf{Known TD} was TD that had been identified (sometime prior) and} was known to the team - either documented as a work item or tacit knowledge). \textbf{Anticipated \textcolor{black}{TD}} \textcolor{black}{was TD that the team anticipated they would create by choosing a particular approach and the decision to incur the \textbf{anticipated TD} was considered as part of planning. \textbf{Unanticipated TD} was either discovered or created during the sprint without having been anticipated or previously discussed\textcolor{black}{, i.e.,} it had not been documented, nor considered during the planning stages.}

Assessment of \textbf{known \textcolor{black}{TD}} and \textbf{anticipated \textcolor{black}{TD}} occurred mainly during fortnightly sprint planning and backlog refinement ceremonies, whereas \textcolor{black}{identification and} assessment of \textcolor{black}{\textbf{unanticipated TD}} occurred during the sprint delivery in either daily stand-ups, after-parties or other deep dive sessions.
\par
When the team assessed \textbf{known \textcolor{black}{TD}} they considered \textcolor{black}{\textit{consequences of inaction}.  For example, the  mistakes that could be made during sideloading.} The team also considered \textit{urgency} \textcolor{black}{expressly}, specifically how urgent it would be to treat the debt to achieve the current release, or to remove links to legacy components that were planned for decommissioning. The team members considered dependencies of other components, and  \textit{urgency} in terms of short term consequences. 
\par
%\faComment \hspace{0cm}\textit{``So I guess I was keen to see some movement on that ``Tech Debt Work item" because there was some I think important work there to completely get off legacy."} (P2-PM)
%\par
%\faComment \hspace{0cm}\textit{``I'm happy to bump this (other ``Tech Debt Work item"). Was there anything in there though, that you would feel you'd want to sort of understand and do to clean up before (P7-DVs work) is released."} (P2-PM)
\faComment \hspace{0cm} P7 (DA - Visualisation): \say{\textit{With regards to that `universal day' (Item identified for removal). We might need to just understand if there has been used elsewhere as well, before we do anything with it. Like we have used date dimensions in everything. So if they're all using this, and I mean if it is legacy, then we'll need to make sure that we address all of that before we go on remove that kind of thing.}}
\vspace{-0.3cm}\\
\par 
 P2 (Manager): \say{\textit{ leaving it there as it is, like, doing nothing. There's no risks in the short term.}}\\ 
%\textit{``Yeah."} (P7-DV)\\ 
\par
\textcolor{black}{In contrast, w}hen assessing \textbf{anticipated \textcolor{black}{TD}}, the team members were observed providing their respective expertise and discussing factors including \textit{familiarity with the type of debt} \textcolor{black}{and the \textit{anticipated impact}} if the debt was incurred, \textcolor{black}{e.g. } \textit{anticipated impact} on system performance:
\par
\faComment \hspace{0cm} P2 (Manager):\textit{``We would be referencing the BIM tables and there's an assessment around the Tech Debt that we would create by doing this piece of work directly off the BIM. Once we do eventually create the DIM layer, we would ideally look to refactor that to be consistent and point to the DIM. Particularly this BIM is more like a FACT table already... It's not the same as a collection of raw tables that we're bringing together for the BIM. P9, I'm looking to you for feedback about what you needed to do previously, was that relatively straightforward to use the BIM directly for that?"}
\vspace{-0.3cm}\\
\par P9 (DA-Backend): \say{\textit{I don't see any difference between the logic that we already developed. The only drawback with this one is that we have to maintain the same filter on this work as well, because we don't have like a single component which filters out criteria, if that makes sense.}}\\
%\textit{``Adding more data to the current one will it cause any performance thing, you know, to the existing table, if we are going to the same BIM?"} (P8-DE)\\
%\textit{``It's unlikely, but yeah, I think because the BIM has already built the linkage. So I think the (bigger) performance impact on this one is (from the previous direct access) which has probably got more layers."} (P9-DE).\\
\par
The Manager summarised and articulated the decision. \textcolor{black}{The Manager considered the \textit{anticipated impact}, the \textit{effort to avoid} and \textit{familiarity with the type of TD} in the context of the sprint goal to proceed with the \textbf{anticipated TD}}:
\par
\faComment \hspace{0cm}  P2 (Manager): \textit{``If we wait to fully develop out the dimensional layer...it would add quite a large dependency or prerequisite, before being able to do our summarized measures. And initially, I'm proposing what we take the same approach with this one. So we've got that sort of equivalent, tech debt on both of those solutions...and take the slightly more tactical approach, particularly because it's just a simple flat table, and it is easy to replumb to a different source later...my appetite is to not take on too much of this within the one sprint."}\\
\par
We also observed decisions about \textbf{unanticipated \textcolor{black}{TD}} during deep-dive sessions, prior to releasing the dashboard. One decision was made under time pressure of release, the other did not have pressure to release. Under this \textit{real time pressure}, the main factor voiced by the Manager in making the decision was \textit{end user usability}. Completion of the work involved additional, but not extensive effort by the analyst team member, but this was not considered a major concern:
\par
\faComment \hspace{0cm} P2 (Manager): \textit{``The next things all really relate to fields names. And this is sort of both what you see in the filter panel, but also when you go to personalize visuals. It may end up being confusing for end users, so that would say to me that we probably should put a prefix or a suffix on these names to make it clear."}
\par
\faComment \hspace{0cm} P12 (DA - Visualisation): \say{\textit{What I've done (since yesterday), ...I have prefixed it ...in the Power BI. We made a backlog card  to fix these(in the VIEW).}}
\vspace{-0.3cm}\\
\par
P2 (Manager): \say{\textit{If that's the pathway we need to go down in the shorter term versus the VIEW, then ok.}}\\
\par
By comparison, where the team made an unanticipated identification of \textbf{\textcolor{black}{technical data component debt}} during a review session, without strong release pressure, the assessment resulted in the decision to refactor of the query (and carry work to the next sprint to ensure the \textcolor{black}{TD} was addressed prior to release:
\par
\faComment \hspace{0cm} P2 (Manager): \textit{``I think we'll need to make a call at the end of the day. And that if there's still a bit of work, there's no outward pressure to get this one released. So if we need a bit more time...I think the important step is to just remove that legacy query."}
\begin{tcolorbox}[enhanced,attach boxed title to top left={yshift=-3mm,yshifttext=-1mm},
  colback=black!5!white,colframe=black!75!black,colbacktitle=black!80!black,
  title=Summary,fonttitle=\bfseries,
  boxed title style={size=small,colframe=black!50!black} ]
 \textit{\textbf{RQ2: How does the team identify and assess technical debt?}}\\
 %\vskip
 \textbf{Identification and Assessment of Technical Debt} is contextual. In the case of \textbf{known \textcolor{black}{TD}}, the team considered the \textit{consequence of inaction} and \textit{urgeny of fixing}. However, if \textbf{unanticipated \textcolor{black}{TD}} is identified\textcolor{black}{, i.e.,}  arises unexpectedly during a sprint, then the team considered the \textit{impact of the \textcolor{black}{TD}} and the \textit{time pressure} of delivery. They distinguished between `real' delivery time pressures compared to `sprint' imposed performance time pressures. The team chose to `carry over' scope to fix \textcolor{black}{TD} when there were no real perceived time pressures. Finally, during planning, the team identified potential or \textbf{anticipated \textcolor{black}{TD}} and considered their \textit{familiarity of the type of debt}, the \textit{effort to avoid} and \textit{anticipated impact} of the \textcolor{black}{TD}.
\end{tcolorbox}

\subsubsection{TD Treatment}
Our observations identified the application and consideration of different types of \textbf{\textcolor{black}{TD} Treatments}, as well as the \textit{evolution} of treatments. The \textbf{\textcolor{black}{TD} Treatments} category includes the \textbf{technical data debt treatments} sub-category\textcolor{black}{, which groups concepts that represent treatments the team used to address \textbf{technical data component debt}, including } \textit{refactoring}, \textit{redevelopment}, and \textit{component removal} concepts. We also identified an associated \textcolor{black}{\textbf{quality assurance of technical data debt treatments} sub-category to conceputalise observations about quality assurance activities carried out by team members when addressing \textbf{technical data component debt}}. This sub-category includes \textcolor{black}{multidisciplinary} concepts including \textit{collaborative testing} and \textit{multidisciplinary reviews} to ensure that the \textbf{technical data debt treatments} are carried out as expected without adverse or unexpected impacts.
\par
When discussing \textit{refactoring}, the team considered updating field names, BIM and Report names, \textit{refactoring} SQL Queries to reference newly created BIMs.
\textit{Component removal} treatment activities included the removal of obsolete SQL notebooks, PowerBI dashboards and PowerBI workspaces no longer required. We identified a relationship between \textit{component removal}, \textit{refactoring} and \textbf{\textcolor{black}{technical data debt treatment quality assurance}} because refactoring activities that touch obsolete components \textcolor{black}{needed} to be completed before \textit{component removal} followed by \textit{collaborative testing}:
\par
\faComment \hspace{0cm} P8 (DA - Backend): \textit{``We identified some refactoring might be required. So in terms of like the actual removal of the BIM and DIMs is probably not complex, but the refactoring before and regression testing after - ensuring that we still produce what we are expecting."}
%\textit{``So (P4-DA) I guess for you confirmation as well...so that cross reference is using some of those original tables?"} (P2-PM)\\
%\textit{``Yes, the cross reference, I think it's referencing the legacy BIM still, I think that needs to be refactored."} (P4-DA)
%\par
%\faComment \hspace{0cm} \textit {``So, (we can remove that load) because we're no longer using that load. We were using a side loaded legacy file, and dependency has been removed now and we're now using the properly ingested summary."} (P2-PM)\\
\par
We observed discussions about \textit{redevelopment} to treat \textcolor{black}{TD}. The reasons for \textit{redevelopment} (for 2 of the components) were that their existing implementations could not be extended to accommodate new requirements. The other \textit{redevelopment} was planned to consolidate a number of different approaches that had been implemented on dashboards to communicate with end-users:\par  
\faComment \hspace{0cm}  P1 (Manager): \textit{``I've asked P3 to raise an item in the backlog, to look for a more effective ways to raise the service messages, that reduce some of the overhead."}
\vspace{-0.3cm}\\
\par 
P2 (Manager): \say{\textit {This work is coming under our continuous improvement area.}}\\
%...We've taken a few different approaches with some of the dashboards (for communicating with end users), there's different levels...So we don't have a real consistency with our production workspaces at this point.}} (P2-PM) \\
\par
The team discussed and was observed performing \textbf{technical data debt quality assurance}, including \textit{collaborative testing} and \textit{multidisciplinary reviews}. Team members incorporated acceptance testing criteria when planning \textbf{technical data debt treatment} work. They performed \textit{collaborative testing} where \textbf{technical data debt treatments} spanned multiple layers of a solution. \textcolor{black}{The team discussed \textit{collaborative}} \textit{testing} after the removal of obsolete \textcolor{black}{data} notebooks. \textcolor{black}{The data analyst that had performed the removal, requested that visualisation analysts refresh} their downstream PowerBI models. \textcolor{black}{The refresh had to be performed in the testing environment,} by team members responsible for data visualisation, to verify that there was no \textcolor{black}{unintended} impact. \textcolor{black}{Each} team member responsible for a PowerBI visualisation performed the tests.  \textcolor{black}{When backend data analysts} \textit{refactored}  BIM/DIMs, \textcolor{black}{they also requested} refreshes and \textcolor{black}{worked with visualisation analysts to} \textit{collaboratively test} the downstream PowerBI model:
\par
%\faComment \hspace{0cm}\textit{``Let's includes extensive regression testing as part of the acceptance criteria for this work to ensure there's no adverse right so maybe add to that a Power BI you know data set refreshing in QAT.} (P1-IM)\\
\faComment \hspace{0cm} P9 (DA - Backend): \say{\textit{I removed all the notebooks and tables or views that's not been used or are inactive....I tested in DEV and it looks all good. I requested to P12 to refresh QA to make sure that there's no impact on the Power BI data set on (the dashboards P12 is responsible for) and it looks okay. Now I need help from P6 to refresh the dashboards (that P6 works on) just to make sure that it's not going to be impacting those two as well. And once that is confirmed then I'll get P11 to review if any further testing is required.}}\\
\par
The team used \textbf{knowledge management strategies} including holding \textit{knowledge sharing} sessions and \textit{documentation} of non-standard practices and implementation. This particular strategy was implemented \textcolor{black}{when one team member resigned from the organisation and the team incorporated knowledge transfer sessions into their sprints to handed over knowledge in several sessions to different} team members: 
\par  
\textit{Knowledge sharing} in the form of creating, or updating confluence documentation and hand-over meetings was used to address gaps in team member knowledge as a result of one of the team members taking up a new role with another organisation. The team member had responsibility for supporting a number of non-standard reports and pipelines and hence the team needed to ensure that they could keep up with the continued support. Whilst the team has some reservations about including this as a `tech debt' related activity, they chose to group it as such for planning and reporting purposes:
\par
\faComment \hspace{0cm} P6 (DA - Visualisation): \textit {``Basically on this one, I'm working on Confluence. One page for each project or dashboard, including the one I am working on at the moment...I am also updating existing documentation which is still, you know, referring to legacy."}\\  
%\textit{``I was calling this one this activity out as a sprint goal. So I've got it under the Tech one. It's not specifically tech debt, but we're putting it under that sort of banner."} (P2-PM)\\
\par
\faComment \hspace{0cm} P2 (Manager): \say{\textit{That's the most sort of critical one, obviously, while you're here you know, capturing as much detail and those knowledge sharing opportunities while you're here.}}\\
\par
We also observed how \textit{organisation structure} facilitated \textit{knowledge sharing}. P1 took on a Management role in another team that provided platform services to the 'Star Squad' and this facilitated \textit{knowledge sharing} during that team's ceremonies. P1 indicated that this was a key factor in facilitating common understanding and improvement of managing \textit{deployment pipeline} debt:
\par
\faComment \hspace{0cm} P1 (Manager): \textit{``It was good to have me now also in the role of Manager with the $<$(other) team$>$ - it helps with the alignment.  We have `common language' when we mention branches and releases... I notice in stand ups, the language seems to just sit. Now, when people talk about `feature branches' we know exactly what they're talking about. \textcolor{black}{So} it seems to have landed with everyone".}\\
\par
\textbf{Environment enhancements} including \textit{process enhancements} were used to address \textit{deployment pipeline} debt. However, the team was unable to fix the technical integration issues relating to the underlying \textit{deployment pipeline}, and instead implemented \textit{process improvements} to address the impact caused by \textit{deployment pipeline}. The team is also actively monitoring availability of \textit{technology enhancements} to address improvements with the PowerBI pipeline:\par
\faComment \hspace{0cm} P11 (Analyst): \textit{``One thing we're doing extra now is a final QA master branch orchestration Power BI refresh. If it works, it's almost guaranteed success. And that's something that has been introduced over the last few weeks. So after our release meeting, we gather up all of the disparate branches from all the teams and we perform a review of the final branch. Then we run the orchestration from start to finish and we do a Power BI refresh. Then we can run a regression test."}\\
%\faComment \hspace{0cm}\textit{``How do we introduce automated regression testing into the release pipeline.  All (regresssion testing) is pretty much all manual. Even if we've got something written for it, we launch it manually. There's no integration of pipelines to run it automatically."}
 \par We identified \textit{evolution} as a relationship of the \textbf{\textcolor{black}{TD} Treatments} category, which provides connection between the treatments available to the team and the wider context of the case. Treatments may \textit{evolve} due to work (carried out by team members and others) as part of data governance activities, Community of Practice activities, or may become available due to wider organisational activities or vendor product releases.  This \textcolor{black}{\textit{evolution}} operates at a different cadence to sprints:\par
\faComment \hspace{0cm} P9 (DA - Backend): \say{\textit{There are multiple Communities of Practice...which run with all team members from the DEG...and we look at naming conventions, BIM and DIM data design practices, visualisation standards....The Community of Practice works at a different speed as our `Star Squad' and involves different people}} \textcolor{black}{(during clarification sessions.}\\ 
\par The \textit{evolution} of \textcolor{black}{TD} treatment is not under the full-control of the team and needs to be evaluated. Once new technology becomes available, the team needs to investigate whether it could be useful and add value:\par
\faComment \hspace{0cm} P7 (DA-Visualisation): \say{\textit{We should look at ``Power BI files and Git repos -  BIP files.}}
\vspace{-0.3cm}\\
\par P1 (Analyst): \say{\textit{They make it easy to track changes (in PowerBI) so you can avoid deploying something you don't want to go.}}\\
\par Finally, we also observed \textbf{avoidance} as a treatment approach for \textcolor{black}{\textbf{anticipated TD}} as the team would try to find ways to de-scope the cause of \textcolor{black}{TD}. For example, old \textbf{data quality debt} from years prior to 2000 would be identified during sprint-planning, and the team took decisions to actively de-scope the data from the feature and \textbf{avoid} incurring the debt. 
\begin{tcolorbox}[enhanced,attach boxed title to top left={yshift=-3mm,yshifttext=-1mm},
  colback=black!5!white,colframe=black!75!black,colbacktitle=black!80!black,
  title=Summary,fonttitle=\bfseries,
  boxed title style={size=small,colframe=black!50!black} ]
  \textbf{\textit{RQ3:What does such as team discuss about technical debt treatment?}}\\
  We identified \textbf{technical data debt treatments} including \textit{refactoring}, \textit{redevelopment} and \textit{component removal} as well as associated \textbf{technical data debt treatment quality assurance}. Multidisciplinary teams rely on \textit{collaborative testing} and \textit{multidisciplinary reviews} where different team members are called to contribute expert knowledge and perform \textbf{technical data debt treatment quality assurance}. We also observed the team discuss efforts to \textbf{avoid} \textcolor{black}{TD},  apply the  \textbf{knowledge management strategies} and make \textbf{technical environment enhancements}. Available treatments \textit{evolve} in line with the team's internal improvement processes, but are also facilitated and influenced by connection with the team's wider context such as COPs, Data Governance and Vendors. External sources of \textit{evolution} do not align with \textcolor{black}{sprint} cadence.% in agile sprints.
\end{tcolorbox}

\subsubsection{Technical Debt Treatment Implementation}\par
We observed several discussions during the team's sprint planning and backlog refinement sessions in which the team defined and refined the way that treatment should be implemented.  We conceptualised these findings into the \textbf{\textcolor{black}{TD} Treatment Implementation} category, which is made up of sub-categories \textbf{define treatment},  \textbf{assess treatment}, and \textbf{\textcolor{black}{TD} treatment work breakdown}.
\par
\textbf{Define treatment} could be as simple as raising one or more backlog tickets during review meetings to capture or identify the \textbf{\textcolor{black}{TD} Treatment}. However, we also observed discussion about the need to allocate time to think through and design appropriate treatment solutions, or even decide whether a treatment was possible:\par 
%\faComment \hspace{0cm}\textit{``Another stream of work is the next steps beyond the the analysis that P4-DA had done around refactoring our source for some of these (metric data) information as well...Rather than us manually uploading that as a CSV file into our platform, like we used to do, we might be able to source some of this from  $<$source system$>$."} (P2-IM) 
\par
\faComment \hspace{0cm} P2 (Manager): \textit{``So we do need to spend a bit of time to really understand the requirements around all of this reference data configuration data. And obviously, I talked about how we managed that in Legacy, which was just in an Excel spreadsheet, where we then exported each worksheet as a separate CSV file to load into the platform. But it'll be worthwhile for someone to have a look through that and propose other options for how we might want to manage that, obviously, it's not ideal managing such critical data in a in a manual form like that, like mistakes can be made. They can have significant impacts to the to the reports and data. A review of what we need to manage in that space will be worthwhile as well."}\\
\par
The \textbf{assess treatment} category has an implied precondition that there is a \textbf{\textcolor{black}{TD} Treatment} available\textcolor{black}{, i.e.,} that \textbf{define treatment} work has been done. We observed the team \textbf{assess treatment} for the purpose of planning implementation. The team discussed the treatment and specifically considered \textit{effort to deliver} the treatment, \textit{complexity of the treatment} and potential \textit{impact to end users}. With respect to \textit{impact on end users}, this included consideration of potential negative and critical impacts of the treatment, in particular the timing given the current point in the business cycle:
\par
\faComment \hspace{0cm} P11 (Analyst): \textit{``And this is the stuff that was considered medium priority last week. Should we just consider like how critical it is to do these? I think (previous technical team member) would have said, `it's quite critical', but P2-PM has a different viewpoint. Especially with these %course related
particular FACT tables, I'll be honest, I don't really want to mess with it midway through this critical
%teaching 
period (for end users) - I'm not saying `Get rid of it'.  But some of those other FACT tables, like, no one uses them, I reckon that's safe."}
\vspace{-0.3cm}\\
\par
\faComment \hspace{0cm} P2 (Manager): \textit{``So if we only implement this spelling change, it will break anyone's existing bookmarks."}\\
\par
We also observed the team \textbf{breaking down \textcolor{black}{TD} treatment work} with the goal of ensuring that the treatment could be delivered within the boundaries of the fortnightly sprint.  When scoping and structuring the work, we observed the team \textit{consider dependencies} and apply \textit{splitting strategies} and \textit{grouping with enhancements}:
\par
\faComment \hspace{0cm} P12 (DA - Visualisation): \textit{``So regarding the ...FACT, we do have the renaming of the FACT included in the refactoring. I'm assuming that renaming is not part of this task, because then that will impact the Power BI report?"}\\ 
\par
The team used \textit{splitting strategies} to structure the work for delivery.  Splitting strategies consider splitting by domain, dependencies, complexity or work impact:\par
%\faComment \hspace{0cm}\textit{``In the previous sprint, we split it (by domain app area) into two activities basically and we have done the work for the first domain. And similarly, we will do the same thing now for the (other domain)."} (P9-DE)
%\par
%\faComment \hspace{0cm}\textit{``This one is pretty much where there's no dependency or there's no active (Azure) workflow that's using  this information so that you can probably safely remove and decouple this. So there are like 1,2,3 or 5 of like objects that's probably involved on that one. And yeah, and it's not been used at the moment."} (P9-DE)
\par
\faComment \hspace{0cm} P8 (DA - Backend): \textit{``Potentially we can split by subject area, for example, like we can do like the DIMs and FACT related to $<$domain area$>$ and the second one will be like the $<$domain area$>$."}
\vspace{-0.3cm}\\
\par
P1 (Manager): \say{\textit{If you look at it, there are 12 there, right? So if we consider there is effort on this from a refactor regression. If you lump it all in one, we could be sitting on it for quite some time."}}\\
%\par
%\faComment \hspace{0cm}\textit{``Do you think you'll do both at the same time? Or are you gonna do low and then push that out? And then the medium afterwards?"} (P11-QA)\\
%\textit{``Yeah, I was thinking to push like the low first, and then the medium probably do it bit later on. So yeah, I just put everything in this card for now and we need to split it out, maybe based on that low and medium complexity."} (P9-DE)

\begin{tcolorbox}[enhanced,attach boxed title to top left={yshift=-3mm,yshifttext=-1mm},
  colback=black!5!white,colframe=black!75!black,colbacktitle=black!80!black,
  title=Summary,fonttitle=\bfseries,
  boxed title style={size=small,colframe=black!50!black} ]
\textit{\textbf{RQ4: How does the team decide the treatment it will apply to the technical debt?}}\\
The team applied steps to select or \textbf{define treatment} and then to \textbf{assess the treatment} in terms of \textit{effort to deliver}, \textit{complexity of treatment delivery} and \textit{impacts to end users}. Treatment implementation tended to be structured to align with sprint goals and timelines and minimise negative or repeated impacts on end-users. To achieve \textbf{\textcolor{black}{TD} treatment work breakdown}, the team applyed \textit{splitting strategies} and \textit{grouping with enhancements}.
\end{tcolorbox}

\section{Discussion, Implications and Recommendations} 
We discuss our key findings about \textcolor{black}{TD} management in a data-intensive software team and triangulate these with related works.  We derive insights about how the team works and make some recommendations for practitioners and researchers that could improve the experience of such multidisciplinary teams and build relevant knowledge for the research community. 

\subsection{Mapping practitioner discussions about TD Types and TD Management to literature}
\subsubsection{TD Types}
Our study provides insight into the \textcolor{black}{TD} types experienced by a single data-intensive software team maintaining and extending a complex enterprise reporting product over a period of 6 weeks. We identified five sub-categories of \textcolor{black}{TD} types including: \textbf{technical data \textcolor{black}{component} debt}, \textbf{data quality debt}, \textbf{pipeline debt}, \textbf{\textcolor{black}{data communication} process debt} and \textbf{\textcolor{black}{legacy} documentation debt}. %We aim to align the identified debt categories to established research and literature to gain insights and determine gaps. 
\textcolor{black}{Based on the} established taxonomy by \cite{Rios2018-yi}, our \textcolor{black}{sub-}categories of %\textbf{process debt},
\textbf{\textcolor{black}{legacy} documentation debt}, \textcolor{black}{\textbf{data communication process debt}} and \textbf{pipeline debt} can be readily aligned with \textcolor{black}{\cite{Rios2018-yi}} categories of Documentation Debt, \textcolor{black}{Process Debt and Infrastructure debt}. 
\par However, we found it somewhat challenging to align the \textbf{technical data component debt} sub-category, and were unable to align the \textbf{data quality debt} sub-category to \cite{Rios2018-yi} TD categories. Rios et al. developed their TD categories through a tertiary mapping study of software engineering projects, \textcolor{black}{which did not specifically consider data-intensive software projects as a type of project and} hence their definitions and example situations of TD items \textcolor{black}{did not focus on the particular nuances of data-intensive software systems.}
\par \textcolor{black}{We considered each of the \textbf{technical data component debt} concepts individually\textcolor{black}{, i.e.,} \textit{layer misalignment}, \textit{naming inconsistency}, \textit{visualisation work arounds} and \textit{obsolete components}). We considered that \textit{layer misalignments} could be aligned as Design Debt \citep{Rios2018-yi}, and \textit{naming inconsistencies, visualisation work arounds} and \textit{obsolete components} could be considered as Code Debt \citep{Rios2018-yi}. Elaboration and inclusion of technical data debt related definitions and examples could be a very helpful extension to the existing Rios et al. TD categories.}

\textcolor{black}{Further,} we note that the \textit{pipeline debt - sideloading} issues have been articulated at a high level by Scully et al.\citep{Sculley2015-hq}, and recently explored by Foidl et al.'s review into pipeline quality and architectures \citep{Foidl2024-tj}. Whilst data model debt type was articulated by Waltersdorfer \citep{Waltersdorfer2020-yi}, their categorisation was limited to data model documentation and we note that further research is required to elaborate and generalise the categories of technical data debt. \par

\subsubsection{TD Management}
\textcolor{black}{We position our findings within the existing TD Management literature, by aligning our findings to Rios et al.'s Technical Debt Management landscape (TDM Landscape), which includes macro activities of Prevention, Identification, Monitoring and Payment of technical debt \citep{Rios2018-yi}. We consider alignment to Perez et al.’s TD prevention strategies \citep{Perez2021-xm}, \textcolor{black}{Freire} et al.’s TD payment map and Payment practices \citep{FREIRE2023} and findings from Li et al.’s case study on SATD \citep{Li2023-zc}.}
\vspace{0.5cm}\\
\textcolor{black}{\textit{5.1.2.1} Contextual Assessment of Technical Debt}
\par \textcolor{black}{Our sub-category of \textbf{known TD} in the category of \textbf{Contextual Assessment of TD} can be readily aligned with the Identification and the Monitoring macro activity groups within the TDM Landscape \citep{Rios2018-yi}. \say{Identification: refers to activities carried out with the objective of identifying debt items} \citep{Rios2018-yi} and \say{Monitoring: refers to monitoring activities of debt items during the evolution of the project} \citep{Rios2018-yi}.}
\par Our other sub-categories in \textbf{Contextual Assessment of TD}: \textbf{anticipated TD} and \textbf{unanticipated TD} represent observations where the team were faced with situations where they had to choose to incur or pay down technical debt. For alignment we consider the concept of Self Admitted Technical Debt (SATD), which \textcolor{black}{\say{is a TD intentionally introduced by developers. It refers to the situation where development teams know that the current implementation is far from ideal and document this by using code comments, alerting the inadequacy of the solution}, \citep{Rios2018-yi} and other artifacts such as commit messages and issue trackers \citep{Li2023-zc}.}

\textcolor{black}{For both \textbf{anticipated TD} and \textbf{unanticipated TD} we consider alignment to Prevention and Payment macro activities in the TDM Landscape \citep{Rios2018-yi}.  \say{Prevention: refers to activities whose objective is to prevent the occurrence of TD} \citep{Rios2018-yi} and \say{Payment: refers to activities undertaken with the goal of supporting decision-making about the most appropriate time to eliminate debt items} \citep{Rios2018-yi}. When \textbf{anticipating TD}, the team considered \textit{effort to avoid}, \textit{the familiarity with the debt} (and the payment required to address it eg. refactoring), and the \textit{anticipated impact} the debt would have (or interest the team would need to pay) on downstream developments. Similar considerations are also reflected in Li et al.’s findings from their interviews with developers about the circumstances in which (SA)TD is ignored and left unresolved \citep{Li2023-zc}, specifically if the repayment of \say{SATD brings small benefit} \citep{Li2023-zc} or \say{repaying SATD takes too much effort} \citep{Li2023-zc}. Developers seem to evaluate the benefit of not incurring TD, suggesting that they do not perceive TD as inherently bad or to be avoided at all costs. The causal relationship between TD prevention and TD payment was explored by \cite{Perez2021-xm}. They identified several causes for incurring TD, including non adoption of good practices and deadlines \citep{Perez2021-xm}. In comparison, the observed team mentioned deadlines (sprint goals) and recognised that they were performing somewhat poor practice, but whilst these may have been underlying causes, they were not deliberated by the team in relation to \textbf{anticipated TD}}.
\par \textcolor{black}{However, when faced with \textbf{unanticipated TD} during the sprint, the team assessed their \textit{time pressure} and also took into consideration the \textit{impact of the technical debt}. This finding aligns with Li et al.'s findings about triggers to pay back SATD, including having “sufficient time” \citep{Li2023-zc}, and whether the (SA)TD is “experienced by stakeholders” \citep{Li2023-zc}}. 
\vspace{0.5cm}\\
\textit{5.1.2.2} Technical Debt Treatments
\par \textcolor{black} {The findings relating to the \textbf{TD Treatment} category can be broadly aligned to the Payment macro activity in the TDM Landscape \citep{Rios2018-yi}, which has been extended by Freire et al.’s Payment map \citep{FREIRE2023}. Freire et al’s Payment Map identifies types of payment related practices relevant to alignment with our \textbf{TD Treatment} category, specifically \say{TD Prevention}, \say{Payment action} and \say{Defining a favourable setting for TP payment} \citep{FREIRE2023}. Individual payment related practices are each aligned with a type of payment related practices \citep{FREIRE2023}}.
\par \textcolor{black}{We align our \textbf{technical data debt treatments} sub-category's \textit{refactoring} and \textit{redevelopment} concepts to Freire et al.’s  \say{Code Refactoring}, and \say{Design Refactoring} payment related practices in the Payment action payment related practice type \citep{FREIRE2023}. However, we could not readily map our sub-category of \textbf{quality assurance of technical data debt treatments}. It is possible that Freire et al. consider quality assurance as  being implicit in the \say{Code Refactoring} \citep{FREIRE2023} payment related practice or that quality assurance in software engineering projects is inherently different to multidisciplinary DI system development. Nevertheless, based on our observations, quality assurance sessions with multiple team members were integral to their work in a multidisciplinary DI team and hence resulted in the creation of its own  sub-category. Interestingly, the performance of\textbf{quality assurance of technical data debt treatments} could possibly also be aligned to Freire et al.'s \say{TD Prevention} type of payment related practices, specifically \say{Code Reviewing} and \say{Investing Effort on Testing Activities} \citep{FREIRE2023}. But in turn, those payment related practices as defined by \textcolor{black}{Freire} et al. lack consideration of multidisciplinary or DI aspects.} 

\par We were unable to align our \textbf{Knowledge management strategies} sub-category at the sub-category level. Our \textit{documentation} concept in the sub-category of \textbf{Knowledge management strategies} can be mapped to Freire et al.’s \say{Update System Documentation} \citep{FREIRE2023} payment related practice, within the \say{Payment action} \citep{FREIRE2023} payment related practice type. But, we could not easily align our related concepts of \textit{organisation structure}, \textit{tacit knowledge} and \textit{knowledge sharing}. They may be likened to \say{Changing the project management} and \say{Improving team collaboration} \citep{FREIRE2023} payment related practices, which are in the \say{Defining a favourable setting for TD payment} \citep{FREIRE2023} type of payment related practices. However, our findings suggest that the observed team relied on \textbf{knowledge management strategies} to address or mitigate the technical debt, not merely to provide a favourable environment to support TD practices. A similar alignment assessment applies to the \textbf{avoidance} sub-category, which we considered for alignment  with the \say{Changing the project scope} \citep{FREIRE2023}, payment related practice, which is also aligned to the \say{Defining a favourable setting for TD payment} \citep{FREIRE2023} type of payment related practices. However, as observed, \textbf{avoidance} via changing the project scope was used to avoid incurring the TD, which may intuitively align to \say{TD Prevention} \citep{FREIRE2023} type payment practices.
   
\par We found it difficult to align the sub-category of \textbf{Environmental enhancements}, which includes \textcolor{black}{the concepts \textit{process improvements}} and \textit{technology enhancements}. Individual concepts within \textbf{Environmental enhancements} such as \textit{technology enhancements} could be considered for alignment to \say{Solving technical issues} \citep{FREIRE2023} related practice in the \say{Payment action} \citep{FREIRE2023} type of payment related practices \citep{FREIRE2023}. Whereas \textit{process improvements} could be considered for alignment to \say{Defining a favourable setting for TD payment} \citep{FREIRE2023} type of activity related practices \citep{FREIRE2023}. We were unable to make an appropriate alignment for the concept of \textit{evolution}, which provides a connection between the team and organisational practices and represents the continuing development of practices and design standards to be applied\textcolor{black}{, i.e.,} development of patterns, design standards and updates to the environment.  
\vspace{0.5cm}\\
\textcolor{black}{\textit{5.1.2.3} Technical Debt Treatment Implementation}
\par \textcolor{black}{Our \textbf{TD Treatment Implementation} category partially aligns to a combination of the Payment and the Monitoring macro activity groups in the TDM Landscape \citep{Rios2018-yi}. The sub-categories of \textbf{define/select treatment} and \textbf{assess treatment} can be aligned to the Payment activity group \citep{Rios2018-yi} and Monitoring activity group (specifically prioritisation and measurement activities) \citep{Rios2018-yi}. The sub-categories also align to \textcolor{black}{Freire} et al.'s TD Payment map, specifically the \say{TD prioritisation} type of payment related practices \citep{FREIRE2023}. We were unable to find appropriate mapping for the sub-category of \textbf{TD treatment work breakdown}, which covers the more detailed aspects of structuring TD payment implementation work in either Rios et al.'s TDM Landscape or \textcolor{black}{Freire} et al.'s Payment map extension. However, we were able to find alignment to the \textit{grouping strategies} concept in Li et al.'s industry case study, which found that \say{grouping TD with enhancements} \citep{Li2023-zc}, and \say{grouping related TD items} \citep{Li2023-zc} are common practices to efficiently pay back SATD \citep{Li2023-zc}.}

\faArrowCircleRight~ \textbf{Implications and Recommendation \#1:} 
\textcolor{black}{Whilst we are able to align most of our observed TD types to the TD types identified in existing literature, we did not find the mapping straightforward and identified some gaps, mainly because the existing TD categories, definitions and examples are based only on software engineering concepts. \textit{We recommend} further research to extend existing TD taxonomies to make these more relevant for DI systems. The extensions should include \textbf{technical data component debt} related definitions and examples and incorporate \textbf{data quality debt} TD type. Whilst our case study is a valuable source for examples and starting point for such research, due to the limitations of a single case study construct, we recommend additional quantitative and mixed methods research to develop generalised findings and understand trend patterns and causal relationships.}   

\faArrowCircleRight~ \textcolor{black}{\textbf{Implications and Recommendation \#2:}}
\textcolor{black}{We found that the TDM Landscape was somewhat challenging to align our category of \textbf{Contextual Assessment of TD}, specifically the situation where teams make decisions to take on TD\textcolor{black}{, i.e., }\textbf{anticipated TD}. The development of guidance and support for taking on TD may be a possible improvement or extension to the TDM Landscape. Our case study provides some examples and insights into the actual discussions and considerations that take place in a team developing a DI system and our findings align with another recent case study by Li et al.\citep{Li2023-zc}. We echo their \textit{recommendation} to conduct further research with practitioners and multidisciplinary teams into investigating the triggers for repaying and not repaying technical debt. We also recommend further research to create guidelines for taking on TD and extending the TDM Landscape and TD payment maps with these guidelines.}

\par

\subsection{Use of TD and TD Management terminology}
Our findings describe the observations that led us to conceptualise \textcolor{black}{TD} Types, and \textcolor{black}{TD} Management Categories including \textbf{\textcolor{black}{TD} Treatment Work}, \textbf{Identify and Assess \textcolor{black}{TD}} and \textbf{\textcolor{black}{TD} Treatment Work Breakdown}.  Whilst basic \textcolor{black}{TD} terminology was used by team members\textcolor{black}{, i.e., }the use of phrases such as `technical debt', `refactor', and `register of technical debt', we did not observe any use of advanced technical debt concepts e.g.  discussions referring to financial metaphors such as `repayment', `interest', `cost' or `anti-patterns'. There were discussions about future refactoring efforts when the team considered the potential to incur tech debt and ease of `replumbing' components. In general, the work was discussed within the context of %agile 
delivery and as such the language and focus are on the type of technical work and the overall value it will deliver, such as `cleaning up', `removing legacy references', implementing new technology or processes and `making consistent'.  %For more complex technical debt items, e.g. the sideloading challenges, the team was considering design of new solution, rather than referencing `design patterns'.
Our prior research identified that one of the characteristics of a multidisciplinary team is that maintaining communication between team members is typically very strong. Also, there is a lot of effort taken by team members to ensure understanding between team members that have different backgrounds \citep{Graetsch2023}. Whilst use of financial metaphors within the software engineering community may be accepted, our observations of a multidisciplinary DI team with several very experienced team members indicated that this \textcolor{black}{may not be} standard terminology in \textcolor{black}{DI system development}.However, as our study was a single, observational case study and due to the inherent \textcolor{black} {nature} of such a study being limited to its context, we are unable to make a more generalised finding. Recently, Xavier et al. proposed and evaluated a lightweight framework to manage \textcolor{black}{TD} in agile software teams, where they incorporated an initial \textcolor{black}{TD} consensus step to allow the team to discuss and agree on \textcolor{black}{TD} definitions for their particular context \citep{Xavier2023-fs}. \textcolor{black} {This approach may be useful for multidisicplinary DI teams.}\par

\faArrowCircleRight~\textbf{Implications and Recommendation \#3:}  The introduction of \textbf{\textcolor{black}{TD} focused language}, and associated financial metaphors could make it easier for team members to consider the longer term implications of taking on debt, or take advantage of downstream concepts such as anti-patterns to quickly recognise and assess \textcolor{black}{TD} and design-patterns that can offer solutions. However, given the nature and varied backgrounds of multidisciplinary team members, care \textcolor{black}{should} be taken when introducing \textcolor{black}{new} terminology. \textit{We recommend} that researchers \textcolor{black}{conduct further research to measure the level of awareness about \textcolor{black}{TD} terminology and management approaches in multidisicplinary DI teams. We also recommend that} researchers and practitioners \textcolor{black}{adopt research methods to co-design TD concepts and vocabulary to make it easy to understand for team members irrespective of their disciplinary background.} \textcolor{black}{Where an agile, specifically Scrum delivery approach is used,} the concepts and vocabulary \textcolor{black}{should} be related to agile delivery and \textcolor{black}{related to DI development work practices}. A co-design based approach could be used to facilitate the development of this guidance. 
%Essentially the language and focus is operational and focused on improvement, not conceptual.  Framework adoption such as LTD may work in this situation?\\
\par
\subsection{TD identification and explicit documentation}
 We did not observe use of special tooling or techniques regarding \textcolor{black}{TD} identification in our case study. However, we did observe that knowledge about \textbf{known \textcolor{black}{TD}} was held in Jira tickets or attached documentation and significant tacit knowledge was evident during discussion of \textbf{anticipated \textcolor{black}{TD}}. Sometimes the discussions resulted in new backlog item creation, but not consistently. Team members in our study did express intent to document/collate TD\textcolor{black}{, i.e.,} make it explicit and the desire to hold broader discussions about TD\textcolor{black}{, i.e.,} with other teams. The team did not use a formal \textcolor{black}{TD} register (other than the backlog) and we did not observe any reference to agreed structures specifically for \textcolor{black}{TD} user stories.  \par
  The potential structure of TD documentation has been considered in software development focused literature such as Li et al. \citep{Li2015-vm} and the preference to keep a separate TD backlog was discussed by Xavier et al. \citep{Xavier2023-fs}. Whether and how well these approaches could meet the needs of data-intensive software has not been investigated. Authors of software development TD management frameworks do consider the \textcolor{black}{creation and maintenance} of a list of \textcolor{black}{TD} to be a foundational element of these frameworks \citep{Guo2016-hc, Xavier2023-fs}. However, the creation and maintenance have to be balanced against the effort of doing so \citep{Guo2016-hc}. 
\par
\faArrowCircleRight~\textbf{Implications and Recommendation \#4:} 
 The need for a product manager to communicate about \textcolor{black}{TD} more broadly implies that information about identified \textcolor{black}{TD} needs to be shareable with broader audiences and \textcolor{black}{TD} management can have broader stakeholders than the immediate team. \textit{We recommend} that practitioners consider the establishment of a \textcolor{black}{TD} register as part of their practices. Whilst further research is needed to clarify what information to capture for optimal downstream management and exactly what \textcolor{black}{TD} should be the focus - the need to have an explicit list of \textcolor{black}{TD} is foundational and it would reduce the risk of maintaining this knowledge tacitly. \textit{We recommend} that the research community conduct further research studies to develop clear guidelines about what data intensive debt should be registered and what information should be captured.
 \par
\textbf{Knowledge management strategies} such as \textit{documentation} and \textit{knowledge sharing} sessions were also used by the team to address \textbf{(non standard) documentation debt}. Data-intensive software engineering is recognised as a highly knowledge intensive activity \citep{Bruce2021-xi} and there has been considerable research about the use of intelligent techniques to drive identification of software-intensive systems \textcolor{black}{TD} from artifacts, as synthesised by \citep{Albuquerque2023-jg}. This research has not been extended to data-intensive systems and it has also not yet extended to converting tacit knowledge about \textcolor{black}{TD} to explicit knowledge.\par
\faArrowCircleRight~\textbf{Implications and Recommendation \#5:}
\textit{We recommend} that researchers extend intelligent technical-debt identification studies to data-intensive debt and also consider research for converting tacit knowledge from discussions into explicit technical-debt documentation or registrations.

\subsection{Multidisciplinary Data Development Tools support for TD Management}
The team had modern development and collaboration tools (see Section 4.1.3), but \textbf{there was no observed support in the tools for identifying or managing \textcolor{black}{TD}}. As we did not observe individual team members perform their work individually, we did not observe whether the tools provided features to identify localised anti-patterns or remedy \textcolor{black}{TD}. However, we did observe team members perform \textit{multidisciplinary review} sessions where they reviewed dashboards and SQL notebooks manually and identified naming inconsistencies, inconsistent configuration of widgets, and access to legacy components and we assumed that the tools either did not have the features or were not set up to guide team members to proactively flag errors.\par
Automated or context-sensitive tool-based support is highly dependent on research into data-intensive \textcolor{black}{TD} anti-patterns - which is still emerging. Muse et al. have identified SQL access anti-patterns through repository mining based research and planning to evaluate these findings with practitioners \textcolor{black}{ \citep{Muse2022-aa}}.\par 
\faArrowCircleRight~\textbf{Implications and Recommendation\#6:} Data engineering tools have an important role in improving the recognition or warning about potential anti-patterns and inconsistencies and hence, prevention of \textcolor{black}{TD}. \textit{We recommend} that researchers conduct further studies to articulate the appropriate, detailed configuration elements for organisational and team-based consistency rules to be incorporated into the tools and evaluate prototypes of tools. \textit{We further recommend} that when developing the tools to identify \textcolor{black}{TD}, developers keep in mind the communication needs of multidisciplinary teams and their stakeholders so that any information produced is readily understood by a range of stakeholders. \textit{We recommend} that tool developers enhance tools by incorporating features to allow configuration of organisational and team based data intensive consistency rules.
\par
When team members were observed as they performed \textbf{multidisciplinary reviews}, one expert team member - usually the team member that had created or updated the SQL or PowerBI Dashboard, would `drive' the tool in response to questions from other team members. There were no features within the tools to capture \textcolor{black}{TD} identification, enable collaboration, or incorporate other \textcolor{black}{TD} related feedback.\par  
\faArrowCircleRight~\textbf{Implications and Recommendation \#7:} The collaborative nature of review sessions where team members other than the `developer' reviewed aspects of SQL or visualisation configurations gives rise to the need for additional features in development tools that support these interactions.  \textit{We recommend} that researchers should evaluate whether collaborative review features such as ability to raise actions, questions and make comments that can be tracked by team members.\\
%\par
%There was also considered the need for `integrated data-lineage \textbf{impact analysis}  across the Databricks and PowerBI platforms. The team even contemplated a `one off' exercise to create a `searchable' list. There was recognition that future implementation of a data catalogue solution would offer data-lineage features and would be on the horizon, but that was a longer term, DEG organisation initiative.\par 
%\textbf{Implication and Recommendation \#7:} 
%There is an urgent need to provide solutions `inter-tool' data lineage analysis in the short-term, perhaps through development of short term utilities.  Whilst strategic, long term solutions are planned, short term solutions are needed.\par

\subsection{TD Treatment Work-Breakdown and Structuring}
From a sprint delivery perspective, work items relating primarily to \textcolor{black}{TD} treatment in our case study were assigned to and measured as `continuous improvement work'. For the prior 4 months, continuous improvement work was kept at 8\% to 9\% of sprint capacity, which equates to roughly 1 or 2 Jira tickets per sprint, depending on the story points assigned.

\par The identified TD treatment work which had been captured in stories by individual team members was considered and scrutinised by different team members during Backlog Refinement and Sprint Planning ceremonies. The multidisciplinary perspectives of the team members contributed to different considerations and led to the identification of how components inter-relate and impact. These discussions identified that further refinement, \textit{splitting}, or \textit{grouping} were required to fit within sprint capacity, risk parameters and value delivery parameters. In effect, the team structured the work to fit into the parameters of the sprint. Whilst the action of splitting and restructuring the debt can be related to  prioritisation of \textcolor{black}{TD}, which has been studied both in the context of software-engineering \textcolor{black}{TD} management \citep{Besker2019-ak} and data-related \textcolor{black}{TD} \citep{Albarak2018-jy}, prioritisation concepts and regimes only consider how to prioritise identified technical-debt.  They do not address how to split and how to structure work, and are hence insufficient to support these activities.
\par \textcolor{black}{Li et al.’s case study made findings that can be related to our findings about structuring of \textbf{TD Treatment Work-Breakdown} for efficiency.  They identified that \say{grouping related technical debt items} \citep{Li2023-zc} or \say{grouping technical debt and development tasks} \citep{Li2023-zc} were efficient practices to pay back SATD \citep{Li2023-zc}. Li et. al also identified a trigger to pay back (SA)TD that could be related to \textit{grouping with enhancements}, specifically \say{SATD is involved in upcoming changes} \citep{Li2023-zc}. They also identified triggers for repaying SATD that could be related to \textit{consider dependencies}, such as if the  \say{SATD hinders other tasks} \citep{Li2023-zc} and \say{the same SATD keeps annoying developers} \citep{Li2023-zc}.  Li et al.’s study did not identify \textit{splitting strategies} as a practice.  However, they did find situations where SATD was ignored or left unresolved including \say{Repaying SATD takes too much effort} \citep{Li2023-zc} and \say{Potential risk of paying back SATD} \citep{Li2023-zc}. It is possible that our observations about the team using splitting strategies represent efforts to address these types of situations. }

\faArrowCircleRight~\textbf{Implications and Recommendation \#8:} 
 \textcolor{black}{Given Li et al.’s case study was not specific to Data Intensive systems and there is alignment with our case study, our findings may potentially be relevant to the broader TD context.} \textit{We recommend} further research to \textcolor{black}{investigate this} and to develop approaches and patterns to break down or structure \textcolor{black}{TD} treatment work \textcolor{black}{ in DI systems and also software systems in general}. This could include staged implementations, or ‘refinance strategies’ that replace highly risky TD, with lower grade TD. The aim is to provide options that can be accommodated \textcolor{black}{within} sprint capacity and \textcolor{black}{align TD Payment to continuous improvement}.  

%\section{Limitations and Threats to Validity} 
\section{Limitations of the Study}
%\subsection{Limitations}
% \textcolor{black}{\subsection{Context}}
% \subsubsection{Findings are limited to the study context}
The context of our observational study is limited to a data intensive development team  working using Scrum for a large organisation and as such we are unable to make claims about generalisation of results. Instead, this case study has aimed to explore details of multidisciplinary DI team practices and build knowledge to support contextualising work and we have provided significant contextual details to meet this aim.

Given that the team used Scrum, we attended and observed the team ceremonies and through these ceremonies identified additional meetings to attend. Our observations spanned many team interactions but not all. There were a number of meetings held that the team or team members attended outside their dedicated ceremonies with additional stakeholders and other teams and as such we need to recognise that our observations were incomplete and could have missed factors that shaped the discussions but were not overtly discussed. Whilst additional observations could have provided more data for analysis, the requirements to obtain voluntary informed consent from all attendees made such observations impractical.

\par Further, we utilised a number of data sources - recorded observations (including Jira card extracts, documentation extracts and tool screenshots), clarification discussions and team Slack messages. However, the team members also communicated via dedicated Slack messages, direct Zoom calls with each other, and we did not have access to these.

\par The study was for a fixed duration of 6 weeks, starting in February 2024, and the results of the study may have been different if we had a longer period of time or picked a different month.  This time was agreed to enable us to see several sprint cycles and ensure that we could observe a number of challenges end to end. Longer observations would have provided us with more data to analyse but it would have potentially made it more difficult to obtain consent from the participants. The nature of this shorter observation period also means we are unable to use our data to establish underlying long term trends in the team. 

\par We also acknowledge contextual factors that could have impacted the work practices, specifically the team having gone through a restructure prior to the commencement of the observations and these work patterns may have impacted technical debt identification and management related conversations.
% STGT method is an inductive analysis method that is grounded in the available data. 
\vspace{0.5cm}
\par STGT can be applied in a limited capacity for basic data analysis to enable the development of concepts and categories without having to develop a theory \citep[Chapter 10]{Hoda24}. A socio-technical grounded theory could potentially provide more depth of concepts resulting in rounding out of categories and their relationships (including possible causal analysis).  Since this was a relatively short observational study and we did not proceed to the advanced stages of theory development, including theoretical saturation, which demonstrates completion of data collection when no new insights are being gained from additional data -- we do not make claims about developing a theory. Future research can extend our work toward a theory.

\par The application of STGT for data analysis should be evaluated through demonstrating credibility and rigour \citep[Chapter 13]{Hoda24}. A key threat to credibility is that the first author, who is an experienced industry practitioner and student, performed all the observations and coding of the data. To mitigate this threat and maintain credibility the analysis approach was reviewed by all authors and reviewed in detail by the 2nd and 3rd authors. The authors reviewed and discussed the approach and reviewed data preparation, filtering, coding, concept and categories development throughout the analysis. Disagreements were resolved through consensus reached through discussion.

\par We demonstrated rigour by presenting examples of memos and conceptualisation. The descriptions of the categories include pertinent examples and evidence in the form of quotes. We also demonstrate the relevance of our findings. As outlined in Section 3.4.3, we conducted a member check process where we made materials available to team members at conclusion of the observations by providing a draft technical report with findings to the project team members and organisation contact and sought their feedback. Five participants attended these presentations. The team members who elected to participate in the member check found the materials representative of their work and insightful.
%\subsection{Threats to Validity}
%\vspace{-0.8cm}
%\subsubsection{Repeatability}
Our results are supported through the design, documentation and application of our case study protocol and associated data management process. Whilst the data collection was carried out by a single researcher, it followed protocol and the protocol documentation was reviewed by the project team and is made available in supplementary materials\footnote{ https://doi.org/10.5281/zenodo.13377355} and could support the repeatability of the case study in a different context.  As the case study was conducted within a ‘live’ environment, an identical study cannot be repeated. Further, due to privacy and commercial confidentiality, we are unable to make the data set available for repeat analysis.\\

%\subsubsection{Credibility and Rigour}

%\subsubsection{Hawthorne effect}
Given the use of an observation study, we acknowledge the threat of  the “Hawthorne effect” \citep{mayo2004human}, which refers to the threat that participants being observed change their behaviour because they are aware of the observations \citep{mayo2004human}. Furthermore, a recent study has found that expert participants performed worse under observation than when not observed \citep{TangH2024} and these effects need to be acknowledged. As part of the informed consent process, each participant was made aware that they would be observed and that observation sessions would be recorded. Prior to each observation session, the first author made participants aware that the session would be recorded and asked for permission - hence ensuring that participants had awareness of the recording. Even though the participants were assured that they would be anonymised, there is a threat that participants did not speak freely during the observations or clarifying interviews. Even though the participants are anonymised for readers who did not partake in the study, they are still likely to be able to identify each other and attribute quotes to each other from the materials presented in the case study. To mitigate this threat, apart from making the statement at the beginning of observations and requests for clarification sessions, the observer did not interrupt the flow of work or request that participants make any adjustments to their work. Also, the participants were already familiar with having their Sprint Review ceremonies recorded and shared with a wider organisational audience.

%\subsection{Bias}
We also need to acknowledge the self-selection bias. The participants expressed curiosity about research and were interested in participating in the study and felt comfortable with the observation protocol - a necessity for obtaining voluntary informed consent. Whilst the first author was able to utilise her extensive experience in project delivery to conduct and analyse the observations, this poses the risk of possible bias. This risk was mitigated through discussions and reviews of codes, concepts, and categories with the other authors and through member checking.\\

\section{Conclusion} 
We designed and conducted a 6-week observational case study of a 12 team member multidisciplinary team that was developing a data analytics system.  We present detailed contextual descriptions of the case context including organisational structures, product and stakeholder information, technology architecture and tools, team member backgrounds, as well as a summary of how the team works in terms of its ceremonies, the work they perform and how the team's work is measured from an organisational perspective. We applied the Socio-Technical Grounded Theory (STGT) \textcolor{black}{\textit{for data analysis}} to develop categories and concepts to categorise types of observed \textcolor{black}{TD} and \textcolor{black}{TD} management. Our findings provide a novel perspective on how a multidisciplinary Data-Intensive software team manages \textcolor{black}{TD}.\par
We identified \textcolor{black}{TD} types that the team deals with, including technical data debt, pipeline debt, process debt, data quality debt and (non) standard document debt.
We conceptualised how the team manages \textcolor{black}{TD}, including through identification and assessment of \textcolor{black}{TD}, \textcolor{black}{TD} work, and \textcolor{black}{TD} treatment work breakdown. By taking this approach, we conceptualised that the team assesses known, anticipated and unanticipated debt differently. We identified \textcolor{black}{TD} data treatments and related treatment quality assurance activities, which are part of \textcolor{black}{TD} treatment work, a broader category than traditional debt repayment. \textcolor{black}{TD} treatments which includes techniques to avoid \textcolor{black}{TD}, knowledge management strategies and environment enhancements. \textcolor{black}{TD} treatments work evolves and can be  influenced by factors outside the team's immediate control, such as organisational communities of practice and technology vendors.  
We also identified insights into how multidisciplinary perspectives of the team members contribute to grouping and splitting of \textcolor{black}{TD} treatment so that it can fit within sprint parameters.\par
Our findings and consideration of the literature raise important implications for the \textcolor{black}{TD} language used by teams, how \textcolor{black}{TD} should be documented, tool requirements and support requirements for structuring \textcolor{black}{TD} treatment.  

\section* {Acknowledgements}
We acknowledge and thank the organisation executive leadership and especially the participant team members for being open and generous with access to their working lives. We also thank the stakeholders who participated in ceremonies for granting us permission to conduct our observation study.\par
The first author would like to acknowledge Prof. Yvonne Dittrich for the time taken to discuss and share insights on practice theory, conducting observations and data analysis.
\par
Graetsch was supported by a Faculty of IT PhD scholarship. Grundy is supported by ARC Laureate Fellowship FL190100035.

\bibliographystyle{elsarticle-harv} 
\bibliography{JSSRevision2_MGRQ2BIB}



%\begin{thebibliography}{00}

%%\bibitem[author(year)]{label}

%\bibitem[Lamport(1994)]{lamport94}
%  \textit{\LaTeX: a document preparation system},
%  Addison Wesley, Massachusetts,

%\end{thebibliography}

\end{document}